\documentclass[aps,pre,reprint,longbibliography,floatfix,superscriptaddress]{revtex4-2}
\usepackage{graphicx}
\usepackage{amsmath,amssymb,amsfonts}
\usepackage{hyperref}
\begin{document}

\title[Two-stage RSA of discorectangles and disks]
{Two-stage random sequential adsorption of discorectangles and disks on a two-dimensional surface}
\author{Nikolai Lebovka}
\email[Corresponding author: ]{lebovka@gmail.com}
\affiliation{Laboratory of Physical Chemistry of Disperse Minerals, F. D. Ovcharenko Institute of Biocolloidal Chemistry, NAS of Ukraine, Kyiv 03142, Ukraine}

\author{Mykhaylo Petryk}
\email{petrykmr@gmail.com}
\affiliation{Ternopil Ivan Puluj National Technical University,56, Ruska Street, Ternopil 46001, Ukraine}

\author{Mykhailo O. Tatochenko}
\email{tatochenkomihail@gmail.com}
\affiliation{Laboratory of Physical Chemistry of Disperse Minerals, F. D. Ovcharenko Institute of Biocolloidal Chemistry, NAS of Ukraine, Kyiv 03142, Ukraine}

\author{Nikolai V. Vygornitskii}
\email{vygornv@gmail.com}
\affiliation{Laboratory of Physical Chemistry of Disperse Minerals, F. D. Ovcharenko Institute of Biocolloidal Chemistry, NAS of Ukraine, Kyiv 03142, Ukraine}

\date{\today}

\begin{abstract}
The different variants of two-stage random sequential adsorption (RSA) models for packing of disks and discorectangles on a two-dimensional (2D) surface 
were investigated. In the SD model, the discorectangles were first deposited and then the disks were added. In the DS model, the disks were first deposited and then discorectangles were added. At the first stage the particles were deposited up to the selected concentration and at the final (second) stage the particles were deposited up to the saturated (jamming) state. The main parameters of the models were the concentration of particles deposited at the first stage, aspect ratio of the discorectangles $\varepsilon$ (length to diameter of ratio  $\varepsilon=l/d$) and disk diameter $D$. All distances were measured using the value of $d$ as a unit of measurement  of linear dimensions, the disk diameter was varied in the interval $D \in [1-10]$, and the aspect ratio value was varied in the interval $\varepsilon\in [1-50]$. The dependencies of the jamming coverage of particles deposited at the second stage versus the parameters of the models were analyzed. The presence of first deposited particles for both models regulated the maximum possible disk diameter, $D_{max}$ (SD model) or the maximum aspect ratio,  $\varepsilon_{max}$ (DS model). This behavior was explained by the deposition of particles in the second stage into triangular (SD model) or elongated (DS model) pores formed by particles deposited at the first stage. The percolation connectivity of disks (SD model) and discorectangles (DS model) for the particles with a hard core and a soft shell structure was analyzed. The disconnectedness was ensured by overlapping of soft shells. The dependencies of connectivity versus the parameters of SD and DS models were also analyzed.  
\end{abstract}

\maketitle

\section{Introduction\label{sec:intro}}

In recent years, adsorption and random packing's of macromolecules and colloidal particles on two-dimensional (2D) substrates have attracted much research and development attention ~\cite{Kubala2022,Adamczyk2022}. Such systems  demonstrated attractive practical applications in electronic, optical, and magnetic devices. The model of random sequential adsorption (RSA) is frequently used as an efficient tool for investigation of deposition processes. In RSA model the particles are deposited sequentially on a 2D substrate without overlapping each other. In the so-called "jamming limit" the surface coverage reaches the saturation limit $\varphi^J$. 

Different types of random and cooperative sequential adsorption models have been studied
~\cite{Evans1993}. The effects particle shape on structure of packing's have 
attracted great interest ~\cite{Lebovka2020}. 
Continuous RSA problems for particles of various shapes, e.g., for disks ~\cite{Talbot2000,Feder1980}, squares ~\cite{Viot1990,Feder1980}, cubic particles ~\cite{Malmir2016}, rectangles ~\cite{Vigil1989,Vigil1990,Viot1992,Ricci1992,Talbot2000,Kasperek2018}, oriented rectangles ~\cite{Petrone2021}
discorectangles ~\cite{Viot1992,Talbot2000,Haiduk2018,Lebovka2020}, rounded rectangles, isosceles and right triangles ~\cite{Ciesla2020}, ellipses ~\cite{Talbot1989,Sherwood1990,Viot1992,Ricci1992,Talbot2000,Haiduk2018},
hard polygons ~\cite{Zhang2018}, spheroids ~\cite{Morga2022}, and  needles ~\cite{Sherwood1990,Tarjus1991,Viot1992,Ricci1992} were analyzed. For elongated particles, the non-monotonic dependencies of surface coverage $\varphi^J$ versus the aspect ratio $\varepsilon$  (width to length ratio) have been typically observed ~\cite{Lebovka2020}. For example, for completely disordered RSA packing of discorectangles a well-defined maximum $\varphi^J =0.583 \pm 0.004$ (at $\varepsilon_{max}\approx 1.46$) was observed ~\cite{Haiduk2018}. This behavior can be explained by appearance of orientation degrees of freedom and excluded volume effects ~\cite{Chaikin2006}.

The spatially continuous RSA models related to simultaneous deposition of mixtures of particles on 2D planar surface  have been investigated  ~\cite{Talbot1989,Meakin1992,Wagaskar2020,Martins2023}. In early studies the adsorption of mixture of hard disks of greatly differing particle diameters was studied theoretically ~\cite{Talbot1989}. The dependence of the jamming limit of large disks as the function of the ratio of deposition rate constants was estimated.  RSA of disks of different sizes has been also investigated using computer simulations ~\cite{Meakin1992}. The different time dependencies of coverage $\varphi (t)$ were observed for the large and small disks. Simulation studies of RSA of binary mixture of disks at different relative rate constants have been recently performed ~\cite{Wagaskar2020}. The radial distribution function and volume distribution of pores were analyzed. For a given diameter ratio the maximum total jamming coverage was observed at some optimum relative rate constant. In two-species antagonistic RSA lattice model the restriction on occupation the nearest-neighbor sites by opposite species was introduced ~\cite{Martins2023}. For this model interconnected adsorption and percolation behavior was observed.

In previous studies different RSA models have been also applied for investigation of particle adsorption  on the heterogeneous (pre-patterned) substrates. For disk-shaped particles the studies of RSA processes on the square landing cells positioned in a square lattice array revealed different deposit morphologies (lattice-like, locally homogeneous, and locally ordered) ~\cite{Araujo2008}. Effect of disk polydispersity on the RSA processes on a square patterned substrate has been also discussed ~\cite{Araujo2008}. Morphological characteristics of the RSA coverings of disk-shaped particles on a nonuniform substrates was studied ~\cite{Stojiljkovic2015}. A surface heterogeneity was produced by preliminary deposition of landing cells (elongated rectangles). The study revealed interesting dependence between the porosity of deposit and the size, shape, density and in cell orientation. 

Different variants of the extended RSA deposition models with partially precovered surfaces have been discussed in early studies~\cite{Adamczyk1997,Adamczyk1998,Weronski2005}.
The two stage RSA models with consecutive deposition of polydisperse mixtures of spherical particles have been developed ~\cite{Adamczyk1998,Manciu2004}. This approach was applied for deposition of different particles at the first and second stages. Particularly, RSA processes at pre-covered  surfaces and adsorption of bimodal mixtures were discussed ~\cite{Adamczyk1998}. Irreversible adsorption of colloid particles on heterogeneous surfaces has been studied ~\cite{Adamczyk2002}. In this RSA model the preliminary adsorption of small spheres was followed by adsorption of larger particles. Theoretical estimation of the available surface and the jamming coverage in the RSA of a binary mixture of disks has been performed ~\cite{Manciu2004}. 

The effects of electrostatic interaction on RSA deposition on partially covered surfaces were studied ~\cite{Weronski2007,Weronski2007a}. The RSA model has been applied for investigation of the deposition of charged polymer nanoparticles on heterogeneous surfaces bearing negative and positive areas of controlled topography ~\cite{Sadowska2021}. The heterogeneity was formed by preliminary deposition of larger particles. The results revealed interesting dependencies of maximum coverage and the structure of deposits versus the heterogeneity degree. 
The resent works also reviews different RSA models for deposition at heterogeneous, pre-patterned  and partially covered substrates~\cite{Adamczyk2017, Sadowska2021, Adamczyk2022, Kubala2022}. Particularly, the percolation, transport properties and possible applications of these functional films in electronic,  optical, magnetic, and biological devices were intensively discussed. 

However, the two stage RSA problem for mixtures of particles of different sorts (e.g. disks and elongated particles) has not been studied in details before to the best of our knowledge. In this work, different variants of a two stage RSA depositions of disks and discorectangles were investigated. In the SD model, the discorectangles were first deposited to some level of coverage and then the disks were added until the state of jamming. In the DS model, the disks were first deposited and then the discorectangles were added. The effects of different parameters(diameter of disks, aspect ratio of discorectangles, and level of preliminary coverage) on the structure of deposits and percolation connectivity of particles inside deposits were studied.

The rest of the paper is organized as follows. Section II presents the computational technical details, main definitions, and examples of patterns of particle packing’s. Section III presents the main results, and the final Section IV summarizes our conclusions.

\section{Main formulations and computational technique \label{sec:methods}}
\begin{figure}[!htbp]
	\centering
	\includegraphics[width=\columnwidth]{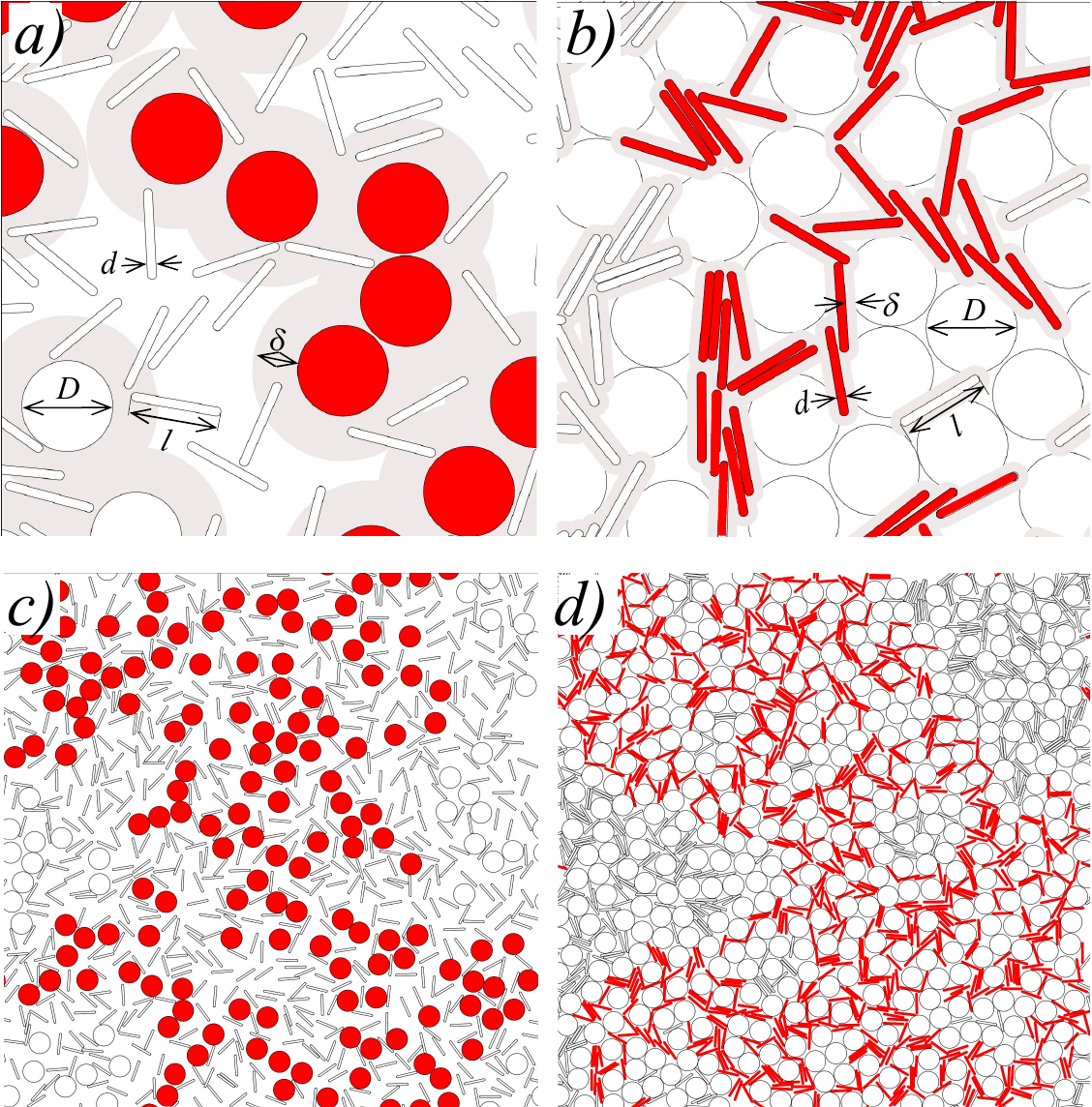}
	\caption{Main definitions for the SD (a) and DS (b) models. Presented patterns are enlarged portions of the size $64\times 64$. Here, $l$ and $d$ are the length and thickness of discorectangle (aspect ratio of was defined as length to diameter of ratio, i.e. $\varepsilon=l/d$), $D$ is a diameter of disk. Connectivity analysis was performed using the particles of the second sort in jamming state (the disks for SD model and discorectangles for DS model). The particles were covered by the shells of thickness of $\delta$. Particles that form a percolation cluster are filled (colored in red) and examples the percolation clusters are presented for SD (c) and DS (d) models. The examples of the patterns are presented for particular cases with parameters: $\varepsilon=10$, $\varphi_\varepsilon^p=0.1$, $D=2$, $\varphi_D^J=0.448$, $\delta=4.96$ (SD model) and for $D=10$, $\varphi_D^p=0.54$, $\varepsilon =10$, $\varphi_\varepsilon^J=0.160$, $\delta=1.08$ (DS model). \label{fig:f01}}
\end{figure}

The adsorption structures were formed using a two-stage RSA model for packing of disks and discorectangles on the 2D plane. At the first stage, 
a preliminary deposition of particle of the first type (disks or discorectangles) was performed, and at the second stage, the particles of another type (discorectangles or disks) were deposited. Two variant of particle deposition were considered (Fig. ~\ref{fig:f01}). In the SD model, the discorectangles were first deposited to some level of coverage $\varphi_\varepsilon^p$ and then the disks were added until they reached their jamming coverage $\varphi_D^J$. In the DS model, the filling procedure of 2D plane was reversed. Here the disks were first deposited to some level of coverage $\varphi_D^p$ and then the discorectangles were added until they reached their jamming coverage $\varphi_\varepsilon^J$. 
An aspect ratio of discorectangles was defined as length to diameter of ratio, i.e. $\varepsilon=l/d$.  Diameter of the disks was defined as $D$. All distances were measured using the value of $d$ as a unit of measurement of linear dimensions. Most of the calculations presented in this paper were performed for intervals $D\in [1-10]$ and $\varepsilon\in [1-50]$. The total size of the systems was $L=L_x=L_y=256$, and periodic boundary conditions were applied along $x$ and $y$ directions.
 
A coverage of the plane by the particles was calculated as $\varphi=NS/L^2$, where $N$ is the number of deposited particles, $S$ is the surface area of  the particle ($S=\pi D^2/4$ for disk and $S=\pi/4+\varepsilon-1$ for discorectangles).
An analysis of the connectedness percolation of RSA packing was always performed for the particles of the second sort in the jamming state, i.e. for the disks in the SD model, and for the discorectangles in the DS model. It was assumed that the particles of the second sort have the hard-core/soft-shell structure with variable thickness of the outer shell $\delta$ (Fig.~\ref{fig:f01}). 
The presence of the outer shell did not affected the RSA process. 

The connectedness percolation procedure was similar to that applied earlier ~\cite{Lebovka2021}. During the connectivity analysis the thickness of the shell was varied and the minimum (critical) value of $\delta$ required for formation of a percolation cluster in the RSA packing was determined.  The analysis was carried out using a list of near-neighbor particles ~\cite{Marck1997} and the calculations were performed using the Hoshen-Kopelman algorithm ~\cite{Hoshen1976}. Particles that form a percolation cluster are filled (colored in red). Figure 1 also presents examples the percolation clusters for SD (c) and DS (d) models (colored in red). 

\begin{figure}[!htbp]
\centering
\includegraphics[width=\columnwidth]{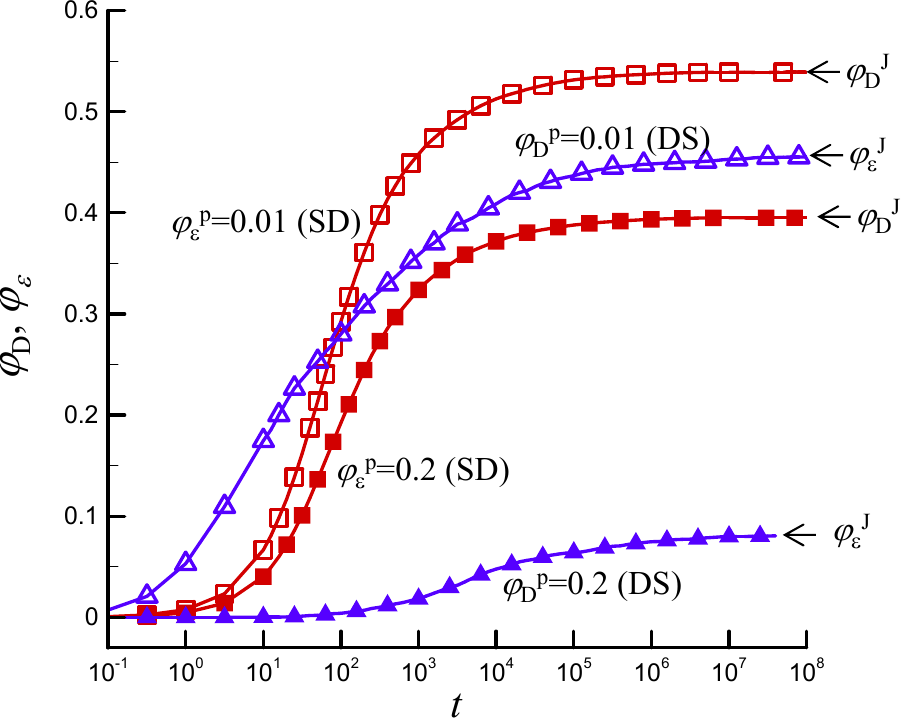}
\caption{Examples of time dependencies of the coverage's during the second stage of the deposition for the SD model (squares) 
$\varphi_D (t)$ and for the DS model (triangles) $\varphi_\varepsilon (t)$. For the SD model the preliminary coverage's by the discorectangles were $\varphi_\varepsilon^p=0.01$ (open squares) and $\varphi_\varepsilon^p =0.2$ (filled squares). For the DS model the preliminary coverage's by the disks were $\varphi_D^p=0.01$ (open triangles) and $\varphi_D^p=0.2$ (filled triangles). The data are presented for $L=256$ and particular cases of $\varepsilon=10$ and $D=1$. Here $\varphi_{D\infty}^J$ and $\varphi_{\varepsilon\infty}^J$ are the jamming coverage’s in the limit of 
$t\rightarrow\infty$ for SD and DS models, respectively.	
\label{fig:f02}}
\end{figure}

Figure ~\ref{fig:f02} presents the examples of time dependencies of the coverage during the second stage of the deposition for the SD model (squares) $\varphi_D(t)$ and for the DS model (triangles) $\varphi_\varepsilon (t)$. For the SD model the preliminary coverage's by the discorectangles were $\varphi_\varepsilon^p =0.01$ (open squares) and $\varphi_\varepsilon^p=0.2$ (filled squares).For the DS model the preliminary coverage's by the disks were $\varphi_D^p=0.01$ (open triangles) and $\varphi_D^p=0.2$ (filled triangles). The data are presented for $L=256$ and particular cases of $\varepsilon =10$ and $D=1$. Here $\varphi_D^J$ and $\varphi_\varepsilon^J$ are the jamming coverage’s in the limit of $t\rightarrow\infty$ for SD and DS models, respectively. 

The deposition time was calculated using dimensionless time units as $t=n/L^2$, where $n$ is the number of deposition attempts ~\cite{Lebovka2021}. The majority of calculations were performed using $L=256$ and the jamming state was typically observed at $t = 10^8-10^{10}$.
\begin{figure}[!htbp]
\centering
\includegraphics[width=\columnwidth]{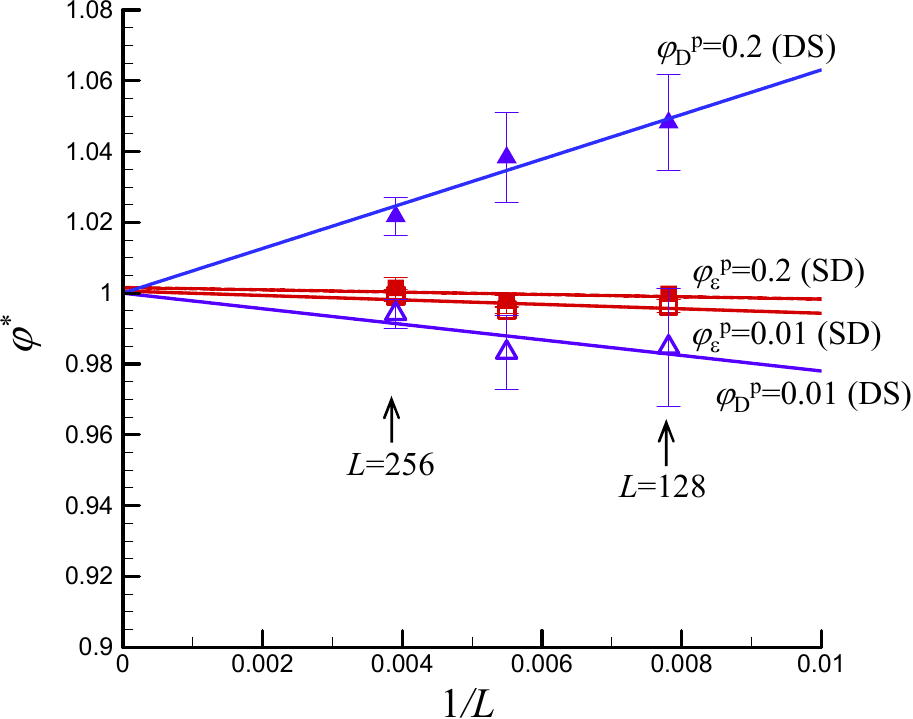}
\caption{Examples of the normalized jamming coverage $\varphi^*$ ($\varphi^*=\varphi_D^J/\varphi_{D\infty}^J$ for the SD model and $\varphi^*=\varphi_\varepsilon^J/\varphi_{\varepsilon\infty}^J$ for the DS model) versus inverse size of the system $1/L$. Here $\varphi_{D\infty}^J$ and $\varphi_{\varepsilon\infty}^J$ are the jamming coverage’s in the limit of $L\rightarrow\infty$. For the SD model the preliminary coverage’s by the discorectangles were $\varphi_\varepsilon^p=0.01$ (open squares) and $\varphi_\varepsilon^p=0.2$ (filled squares), for the DS model the preliminary coverage’s by the disks were $\varphi_D^p=0.01$ (open squares) and $\varphi_D^p=0.2$ (filled squares). The data are presented for $L=256$ and particular cases of $\varepsilon =10$ and $D=1$. 
\label{fig:f03}}
\end{figure}

Figure ~\ref{fig:f03} presents the examples of the normalized jamming coverage $\varphi^*$ 
($\varphi^*=\varphi_D^J/\varphi^J_D{_\infty}$ for the SD model and   
$\varphi^*=\varphi_\varepsilon^J/\varphi^J_\varepsilon{_\infty}$ for the DS model) versus the inverse size of the system $1/L$ for different preliminary coverage’s. The data are presented for $L=256$ and particular cases of $\varepsilon =10$ and $D=1$. 

The jamming coverage’s in the limits of $L\rightarrow\infty$, $\varphi_{D\infty}^J$ (SD model) and $\varphi_{\varepsilon\infty}^J$ (DS model) were estimated assuming linear $\varphi_D^J$ and $\varphi_\varepsilon^J$ versus $1/L$ dependencies.  	

For each given set of parameters, the computer experiments were averaged over 10-100 independent runs. The error bars in the figures correspond to the standard errors of the means. When not shown explicitly, they are of the order of the marker size.

\section{Results and Discussion\label{sec:results}}
\subsection{SD model}
\begin{figure}[!htbp]
	\centering
	\includegraphics[width=\columnwidth]{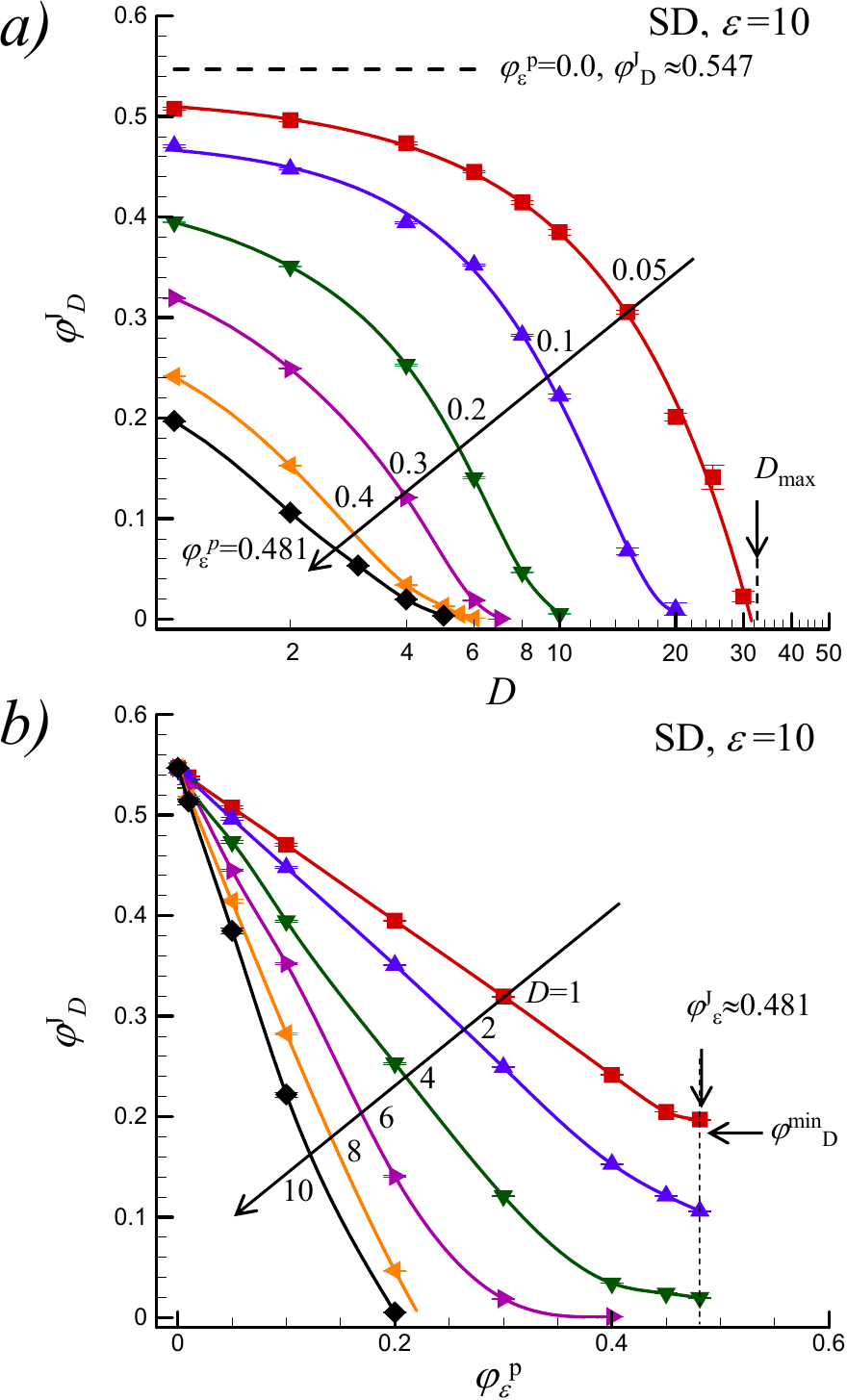}
	\caption{Jamming coverage for disks $\varphi_D^J$ versus their diameters $D$ at different concentration of first deposited discorectangles $\varphi_\varepsilon^p$ (a) and versus $\varphi_\varepsilon$ at different values of $D$ (b). The data are presented for the SD model at fixed aspect ratio $\varepsilon =10$. The values $\varphi_D^J\approx 0.547$ (a) and $\varphi_\varepsilon^J\approx 0.481$(b) are the jamming coverage’s for disks and discorectangles deposited on empty surfaces, respectively. Here, the value of $D_{max}$ corresponds to the limiting (maximum) diameter of the disk (a) and the value of $\varphi^{min}_D$ corresponds to the minimum value of $\varphi_D^J$ at  $\varphi_\varepsilon^J\approx 0.481$ (b).  
		\label{fig:f04}}
\end{figure}

For SD model the discorectangles were first deposited and then the disks were added. Figure ~\ref{fig:f04} presents examples of jamming coverage's $\varphi_D^J$ behavior for disks. Here, the dependencies of $\varphi_D^J$ versus the disk diameter $D$ at fixed values of $\varphi_\varepsilon^p$ (a) and versus concentration of discorectangles $\varphi_\varepsilon^p$ at fixed values of $D$ (b) are given. The value of aspect ratio was fixed at $\varepsilon=10$. The similar dependencies were observed for others values of $\varepsilon$. Preliminary deposition of discorectangles resulted in decreasing of $\varphi_D^J$ (Fig. ~\ref{fig:f04}a). For example, at $\varphi_\varepsilon^p =0.05$ and $D=1$ we have $\varphi_D^J=0.508\pm0.002$ that is noticeably smaller than the jamming limit for the disks on empty surface without the sticks,  $\varphi_D^J\approx 0.547$ ~\cite{Hinrichsen1986}. The value of $\varphi_D^J$ decreased with increasing of $D$ (Fig.~\ref{fig:f04}a) and increasing of $\varphi_\varepsilon$ (Fig. ~\ref{fig:f04}b). 

Obtained data evidenced that above some maximum value of $D_{max}$ the deposition of disks was practically absent (i.e., the probability of deposition was very small). In this work, the value of $D_{max}$ was defined as the maximum value of $D$ at rather small coverage, $\varphi_D^J=0.01$. The value of $D_{max}$ depends upon values of $\varepsilon$ and $\varphi_\varepsilon^p$. For example at $\varepsilon =10$ and $\varphi_\varepsilon^p=0.2$ we have $D_{max}\approx 10$. Note that the value of $\varphi_\varepsilon^p$ can not exceed the jamming coverage of discorectangles at a given $\varepsilon$ (e.g., $\varphi_\varepsilon^J\approx 0.481$ at $\varepsilon=10$, Fig. ~\ref{fig:f04}b). At fixed value of $\varphi_\varepsilon^p$ the value of $\varphi_D^J$ decreased with increasing of $D$. At jamming coverage for first deposited discorectangles, i.e., at $\varphi_\varepsilon^p = \varphi_\varepsilon^J$  (e.g., $\varphi_\varepsilon^J \approx 0.481$ for $\varepsilon =10$ in Fig. ~\ref{fig:f04}b) a minimum value of $\varphi_D^J$($= \varphi^{min}_D)$ was observed. 
The defined above parameters of the maximum diameter of the disk $D_{max}$ and the minimum jamming coverage for the disks $\varphi^{min}_D$ were significantly dependence versus the aspect ratio of first deposited discorectangles $\varepsilon$. 

\begin{figure}[!htbp]
	\centering	
	\includegraphics[width=\columnwidth]{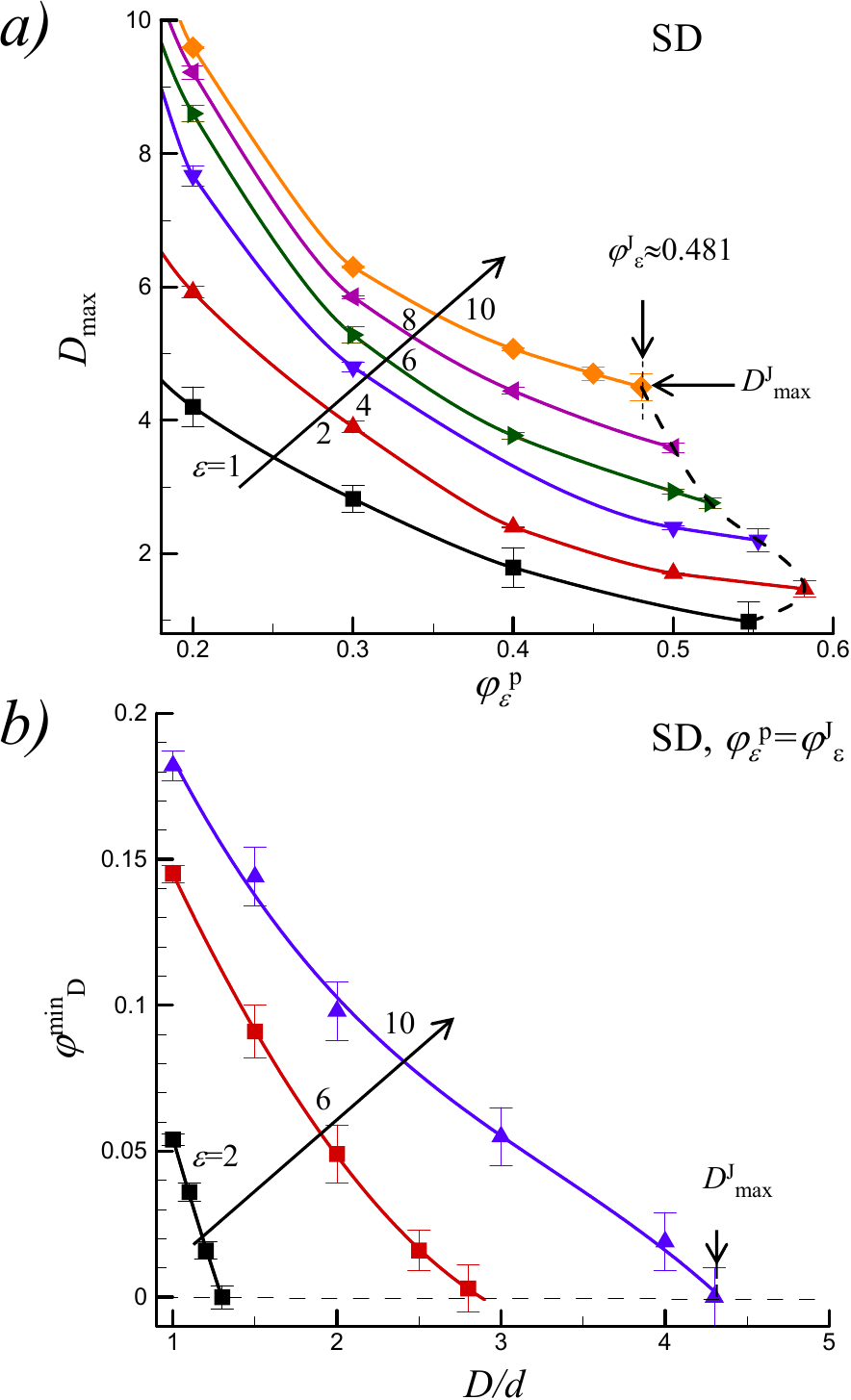}\\
	\caption{Maximum diameter of the disk $D_{max}$ versus the concentration of first deposited discorectangles $\varphi_\varepsilon^p$ (a) and the minimum jamming coverage for the disks $\varphi^{min}_D$ (Fig.  ~\ref{fig:f03}b) versus the $D$ (b). The data are presented for the SD model and different values of aspect ratio of $\varepsilon$. Dashed line in (a) shows values of $\varphi_\varepsilon^J$ in a jamming state. The value $D_{max}^J$ corresponds to the maximum value at $\varphi_\varepsilon^p=\varphi_\varepsilon^J$.
		\label{fig:f05}}
\end{figure}

Figure  ~\ref{fig:f05}a presents $D_{max}$ versus the concentration $\varphi_\varepsilon^p$ at different aspect ratios $\varepsilon$. The value of $D_{max}$ decreased with increasing of $\varphi_\varepsilon^p$ and reached its minimum for jamming coverage of discorectangles $\varphi_\varepsilon^J$ at the given $\varepsilon$. 
Note that the dependence $\varphi_\varepsilon^J (\varepsilon)$ demonstrated well-defined maximum at $\varphi_\varepsilon^J\approx 0.583$ and $\varepsilon \approx 1.46$ ~\cite{Haiduk2018,Lebovka2020}. Figure 5b present $\varphi^{min}_D$ versus $D$ at different values of $\varepsilon$ for preliminary deposition of discorectangles up to the jamming state, $\varphi_\varepsilon^p =\varphi_\varepsilon^J$. The value of $\varphi^{min}_D$ decreased with increasing of $D$ and became zero above some maximum value of $D =D_{max}^J$. 
\begin{figure}[!htbp]
	\centering	
	\includegraphics[width=\columnwidth]{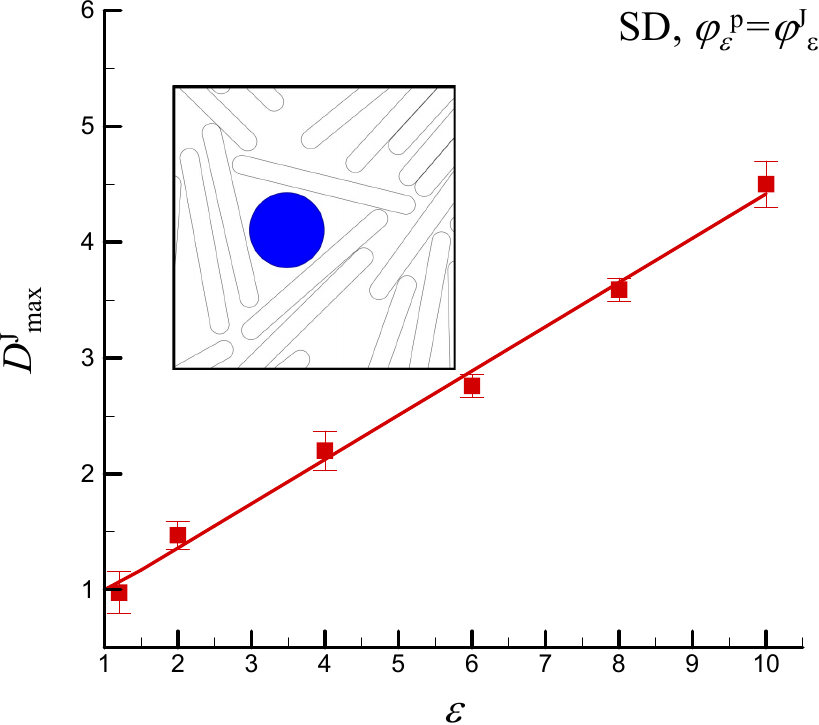}\\
	\caption{Maximum diameter of a disk $D_{max}^J$ versus the aspect ratio $\varepsilon$ of first deposited discorectangles for the fixed their concentration $\varphi_\varepsilon^p=\varphi_\varepsilon^J$  (jamming state). The line corresponds to the linear approximation in Eq. (1). Insert shows the example of packing pattern of size $20\times 20$ for the following parameters: $\varepsilon =10$, $\varphi_\varepsilon^p=\varphi_\varepsilon^J\approx 0.481$, $D=4$. \label{fig:f06}}
\end{figure}
Figure ~\ref{fig:f06} shows the maximum diameter of the disk $D^J_{max}$ versus the aspect ratio of first deposited discorectangles $\varepsilon$ up to the jamming limit with the coverage $\varphi_\varepsilon^p =\varphi_\varepsilon^J$. This dependence can be well approximated by the linear function:
\begin{equation}\label{eq:1}
	D^J_{max}=1+\alpha(\varepsilon -1),
\end{equation}
where $\alpha =0.38 \pm 0.02$.

The linear character of $D^J_{max}(\varepsilon)$ dependence can be explained on the basis of the following simple geometric arguments. In the packing's of first deposited discorectangles the formation of stacks of nearly parallel particles and creation of large “triangular pores” was typically observed. During the second stage of adsorption, the disks can be adsorbed only in such large pores between stacks (see inset in the Fig. ~\ref{fig:f06} with example of the packing pattern). For an ideal equilateral "triangular pore" with side length $\varepsilon$, the diameter of the disk inscribed inside the pore is determined by the formula $D=\gamma\varepsilon$, where $\gamma=1/\sqrt{3}\approx 0.58$. The difference between values of  $\alpha$ and $\gamma$ can reflect nonideality of the “triangular pores” and their smaller sizes in real packing's. 
\begin{figure}[!htbp]
	\centering
	\includegraphics[width=\columnwidth]{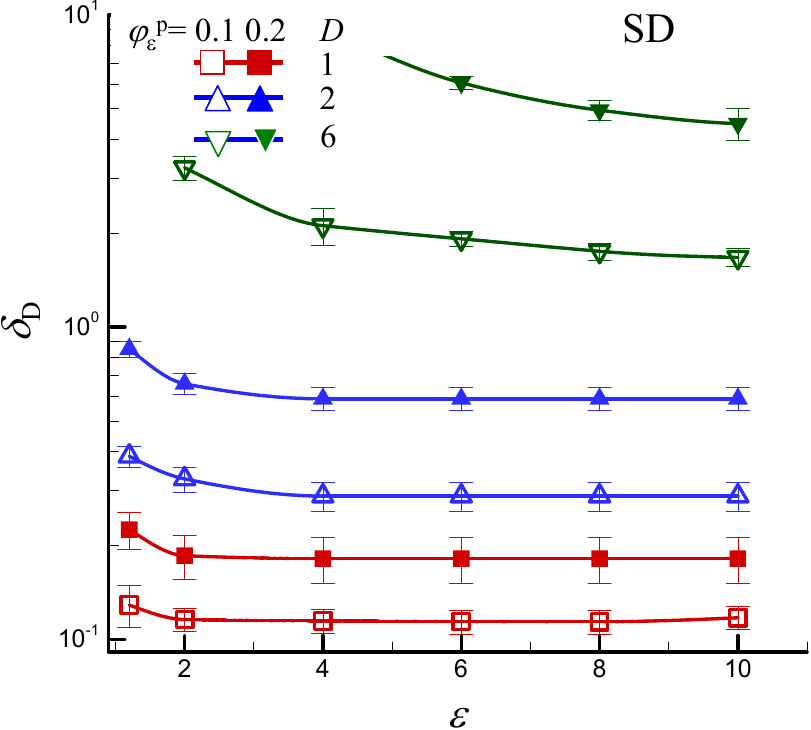}
	\caption{Percolation thickness of a disk shell $\delta_D$ versus the aspect ratio $\varepsilon$ of first deposited discorectangles for their concentrations $\varphi_\varepsilon^p=0.1$ and $\varphi_\varepsilon^p=0.2$, and diameters of the disks  $D=1,2$ and $6$. \label{fig:f07}}
\end{figure}

Figure ~\ref{fig:f07} presents a percolation thickness of the shells around the disks $\delta_D$ versus the aspect ratio of first deposited discorectangles $\varepsilon$. For disks with core-shell structure at this percolation thickness the formation of spanning cluster through the entire system was observed. In particular case of $\varphi_\varepsilon^p=0$ and jamming coverage of plane by disks ($\varphi_D ^J \approx 0.547$) the shell thickness was estimated to be $\delta_D=0.0843 \pm  0.001$. The total coverage of a plane by disks with shells was estimated to be $0.642 \pm 0.001$. Note that this value a little less than estimated total coverage for overlapping disks of equal diameter at the percolation threshold, 
($\varphi \approx 0.676339$) ~\cite{Quintanilla2000}.

The observed behavior for different diameters of the disks $D$ and concentration of discorectangles $\varphi_\varepsilon^p$ (Fig. ~\ref{fig:f07})
can be explained using the following arguments. The preliminary coverage by discorectangles resulted in reducing of probability of deposition of disks at the second stage in near neighbor vicinity to each–others.  This tendency is enhanced with increasing of $\varphi_\varepsilon^p$ and $D$ and both these factors resulted in increasing of $\delta_D$ (Fig. ~\ref{fig:f07}). The weak dependencies of shell thickness at $D=1,2$ may reflect the insignificant impact of first deposited discorectangles at small concentrations $\varphi_\varepsilon^p=0.1, 0.2$
on the connectivity of jammed networks of disks. The significant effects of aspect ratio $\varepsilon$ on the disk connectivity was only observed at relatively large concentration of discorectangles $\varphi_\varepsilon^p$  for commensurate values of  $D$ and $\varepsilon$. Its evidently reflects the separation of disks at large distances with their location in pores between the stacks (see inset to Fig. ~\ref{fig:f06}).
\begin{figure}[!htbp]
	\centering
	\includegraphics[width=\columnwidth]{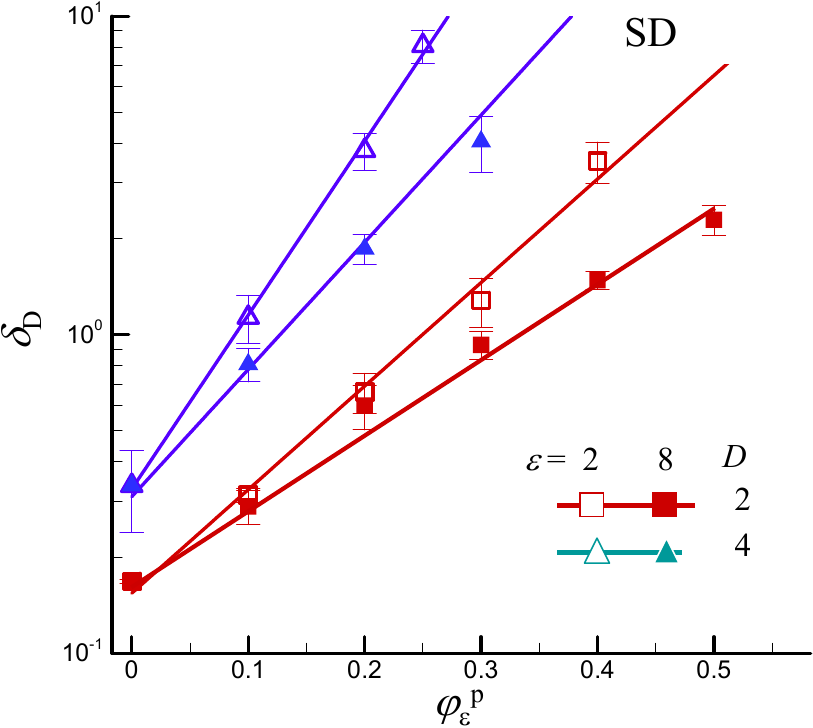}
	\caption{Percolation thickness of the disk shell $\delta_D$ versus the concentration of first deposited discorectangles $\varphi_\varepsilon ^p$. The data are presented for $D= 2, 4$ and $\varepsilon=2, 8$. \label{fig:f08}}
\end{figure}

Figure ~\ref{fig:f08} illustrates examples of $\delta_D$ versus $\varphi_\varepsilon$  dependencies for several values of $D$ and $\varepsilon$. In absence of preliminary deposition of discorectangles (at $\varphi_\varepsilon =0$) the percolation thickness was relatively small and proportional to the disk diameter,  $\delta_D$=$aD$), where  $a=0.084\pm0.001$. However, the $\delta_D (\varphi_\varepsilon)$ dependencies were rather strong (practically exponential) and at large values of $\varphi_\varepsilon$ the percolation thickness of a disk shell $\delta_D$ may significantly exceed the value of $D$.
\begin{figure}[!htbp]
\centering
\includegraphics[width=\columnwidth]{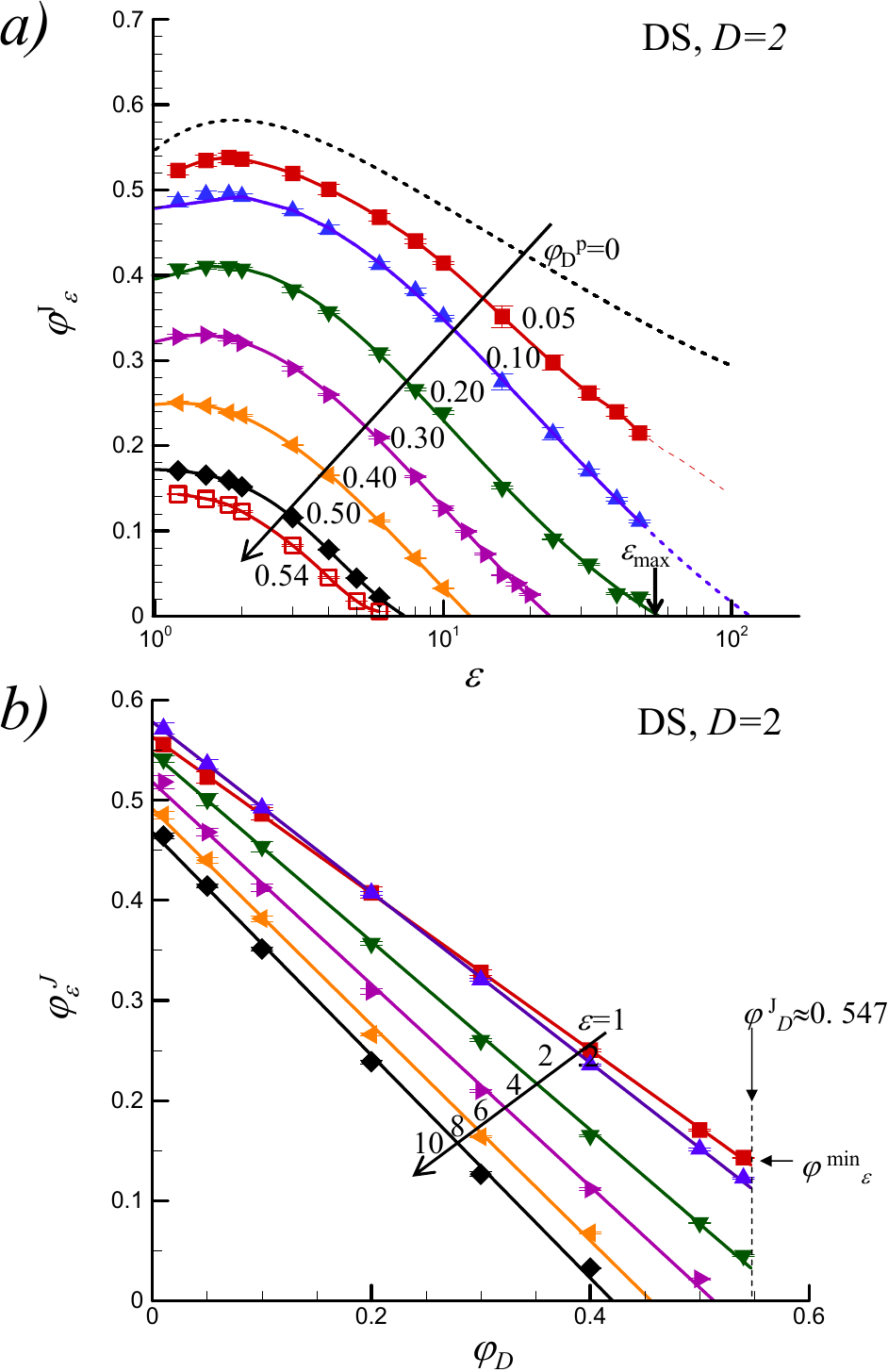}
\caption{Jamming coverage of discorectangles $\varphi_\varepsilon^J$  versus the aspect ratio $\varepsilon$ at different coverage of first deposited disks $\varphi_D^p$ (a) and versus $\varphi_D^p$ at different aspect ratio $\varepsilon$  (b). The data are presented for the DS model at fixed diameter $D=2$. The value $\varepsilon_{max}$ (a) is the maximum aspect ratio of discorectangle that can be deposited for the given value of $\varphi_D^p$. The value $\varphi_\varepsilon^{min}$  is the minimum coverage of discorectangles for the coverage of first deposited disks $\varphi_D^p=\varphi_D^J \approx0.547$ (jamming state).  \label{fig:f09}}
\end{figure}
\begin{figure}[!htbp]
	\centering
	\includegraphics[width=\columnwidth]{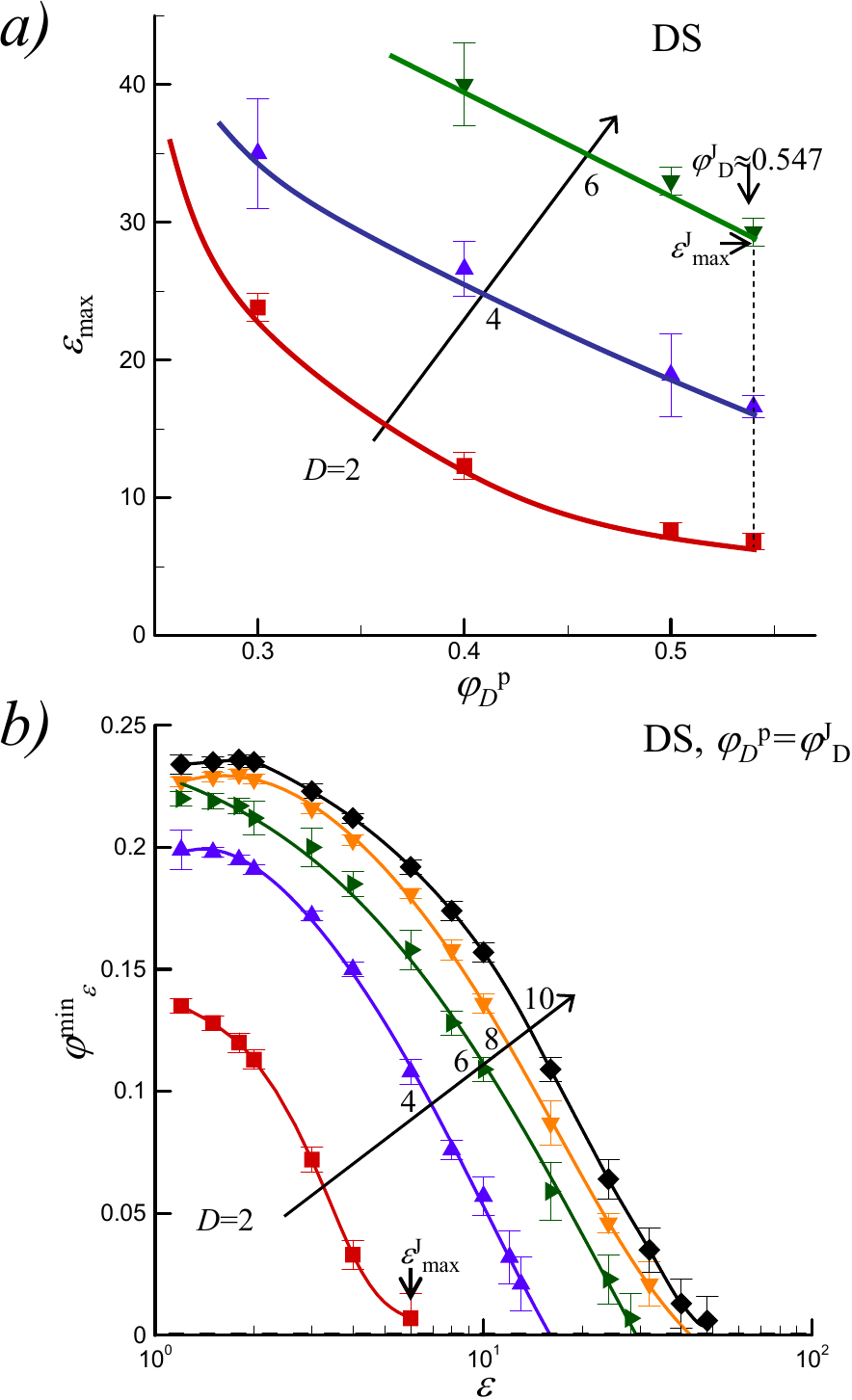}
	\caption{Maximum aspect ratio of discorectangle $\varepsilon_{max}$ (a) versus the coverage of first deposited disks $\varphi_D^p$, and the minimum coverage of discorectangles $\varphi_\varepsilon^{min}$  (for the coverage of first deposited disks $\varphi_D^p=\varphi_D^J \approx0.547$, jamming state) versus the aspect ratio $\varepsilon$  (b). The data are presented for the DS model at at several values of $D$. \label{fig:f10}}
\end{figure}

\subsection{DS model}
For DS model the disks were first deposited and then the discorectangles were added. 
Figure ~\ref{fig:f09} presents examples of jamming coverage's $\varphi_\varepsilon^p$ behavior for discorectangles. Here, the dependencies $\varphi_\varepsilon^J$ versus the aspect ration $\varepsilon$ (a) and versus the concentrations of first deposited disks $\varphi_D^p$ (b) are shown. The data are presented for the fixed $D=2$. For deposition on uncovered surface ($\varphi_D^p=0$) a well-defined  maximum $\varphi_{\varepsilon,m}^J\approx 0.583$ at $\varepsilon\approx 1.46$ was observed ~\cite{Haiduk2018,Lebovka2020}. 

For preliminary covered surfaces the value $\varphi_{\varepsilon,m}$ decreased with increasing of  $\varphi_D^p$, and particularly, at the jamming point $\varphi_D^p= \varphi_D^J\approx 0.547$ we have $\varphi_\varepsilon^J\approx 0.14$ (Fig. ~\ref{fig:f09}a).
Obtained data also evidenced that above some maximum value of 
$\varepsilon_{max}$ the deposition of discorectangles was practically absent (i.e., the probability of their deposition was very small). In this work, the value of $\varepsilon_{max}$ was defined as the maximum value of $\varepsilon$ at $\varphi_\varepsilon^J=0.01$.

The values of $\varphi_\varepsilon^J$ approximately linearly decreased with increasing of the concentrations of first deposited disks $\varphi_D^p$ up to the value $\varphi_\varepsilon^{min}$ at $\varphi_D^p\leqslant \varphi_D^J\approx 0.583$ (Fig. ~\ref{fig:f09}b)

Figure ~\ref{fig:f10} presents $\varepsilon_{max}$ versus $\varphi_D$ (a) and $\varphi_\varepsilon^{min}$ versus $\varepsilon$ (b) dependencies at several values of $D$. The value of $\varepsilon_{max}$  decreased with increasing of  $\varphi_D$ up to the minimum value at $\varphi_D^J\approx 0.547$ (jamming state for first deposited disks). Otherwise, at fixed value of $\varphi_D$ the value of $\varepsilon_{max}$  increased with increasing of $D$ (~\ref{fig:f10}a). 
\begin{figure}[!htbp]
	\centering
	\includegraphics[width=\columnwidth]{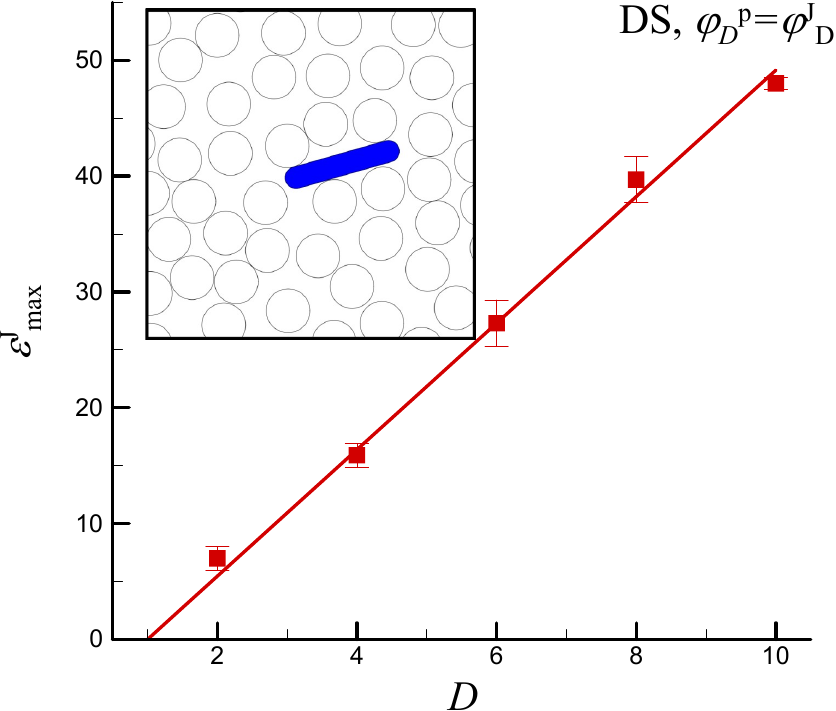}
	\caption{Maximum aspe{\normalsize {\Large }}ct ratio of discorectangle $\varepsilon_{max}^J$ versus the relative diameter of first deposited disks $D$ for the fixed concentration $\varphi_D^p=\varphi_D^J\approx 0.547$ (at the jamming state). The line corresponds to the linear approximation in Eq. (2). Insert shows the example of packing pattern of size $20\times 20$ for the following parameters: $D=2$, $\varepsilon=5.4$. \label{fig:f11}}
\end{figure}
This behavior may be explained by formation of more large pores suitable for deposition of discorectangles at large $D$. The value $\varphi_\varepsilon^{min}$ (at $\varphi_D^p\leqslant \varphi_D^J\approx 0.583$) decreased up to the zero at $\varepsilon =\varepsilon_{max}$. with increasing of $\varepsilon$ (Fig. ~\ref{fig:f10}b). Moreover, the value $\varepsilon_{max}$ increased with increasing of $D$.   

Figure ~\ref{fig:f11} presents the maximum aspect ratio of the discorectangle $\varepsilon_{max}$ versus the diameter od first deposited disks $D$. This dependence can be well approximated by the linear function:
\begin{equation}\label{eq:2}
	\varepsilon_{max}^J =\beta (D-1),
\end{equation}
where $\beta=5.46\pm 0.26$.

The inset to the Fig. ~\ref{fig:f11} demonstrate the example of RSA parking for the DS model with preliminary parking of disks at $D=2$, $\varphi_D^p =0.54$ (close to the jamming state), and one discorectangle with aspect ratio $\varepsilon=5.4$ (close to the value $\varepsilon_{max}^J$). It can be clearly seen that the value of $\varepsilon_{max}^J$ is defined by the dimensions of “elongated” pores inside the preliminary parking of disks. 

\begin{figure}[!htbp]
	\centering
	\includegraphics[width=\columnwidth]{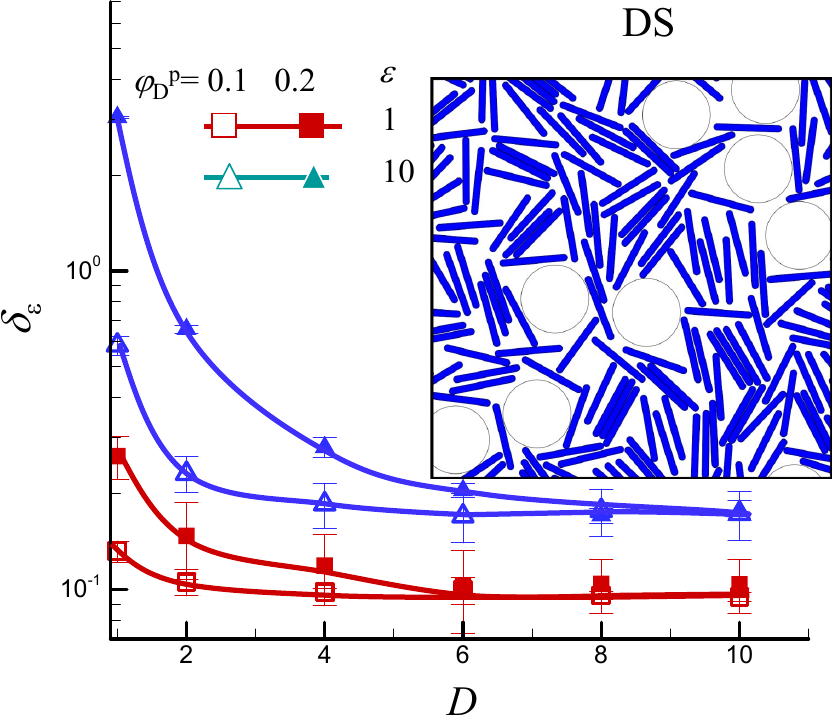}
	\caption{Percolation thickness of discorectangle shell $\delta_\varepsilon$ versus the diameter of first deposited disks $D$ for their concentrations $\varphi_D^p =0.1$, $\varphi_D^p=0.2$, and aspect ratios $\varepsilon =1$ and $10$. Insert shows the example of packing pattern of size $64x64$ for the following parameters: $D=10$, $\varphi_D^p=0.2$, and $\varepsilon =10$.  \label{fig:f12}}
\end{figure}
Figure ~\ref{fig:f12} presents a percolation thickness of the shells around the discorectangles $\delta_\varepsilon$ versus the diameter of first deposited disks $D$. For discorectangles with core-shell structure at this percolation thickness the formation of spanning cluster through the entire system was observed. The value of $\delta_\varepsilon$ decreased up to some asymptotic value with increasing of $D$ and increased with increasing $\varepsilon$. Such behavior can by explained by the following arguments. At relatively large $D$ the first deposited disks can be considered as large inclusions in the packaging of the discorectangles (see inset to the Fig. ~\ref{fig:f07}). In this case the connectivity of the the discorectangles can be only determined by the value of $\varepsilon$.

\section{Conclusion\label{sec:conclusion}}
A study of the two-stage RSA packing of discorectangles and disks on a plane surface was carried out. Two models were analyzed. In the SD model, the discorectangles were first deposited, and then disks were added. The situation was reversed in the DS model. Here the disks were preliminary and then discorectangles were added. For both deposition models the presence of first deposited particles significantly affected the properties of packings formed at the second stage. Particularly for jamming packing formed at the first stage there were observed the limiting maximum values of disk diameter $D^J_{max}$  (model SD) or aspect ratio $\varepsilon_{max}^J$ (model DS). Moreover, the linearly proportional dependencies of type $D^J_{max}\propto \varepsilon $ (model SD) and $\varepsilon_{max}^J\propto D$ (model DS) were observed in both cases. It is interesting that at relatively small preliminary coverages the near linear $\varphi_D^J$ versus $\varphi_\varepsilon^p$ (SD model) and $\varphi_\varepsilon^J$ versus $\varphi_D^p$ (DS model) decreasing dependencies were observed. Such behavior may reflect the specific impact of preliminary deposited particles at the first stage on the jamming coverage of particles deposited at the second stage. 

Using the hard core - soft shell particle model the percolation connectivity of the particles deposited at the second stage was analyzed. For the SD model the percolation shell thickness $\delta_D$ decreased with increasing of both values $\varepsilon$ and $D$. The value of $\delta_D$ exponentially increased with increasing the concentration of first deposited discorectangles $\varphi_\varepsilon$. For the DS model the percolation shell thickness $\delta_\varepsilon$ decreased with increasing of $D$ and increased with increasing of both the concentration of first deposited disks $\varphi_D$ and aspect ratio $\varepsilon$. Such behavior evidence the possibility of fine regulation of the connectivity and transport behavior in films obtained by two-stage adsorption procedure. In further studies it is desirable to consider simultaneous RSA co-deposition of mixtures of particles with different shapes and evaluation of percolation and transport properties of such multicomponent films.

\begin{acknowledgments}
We acknowledge the funding from NASU (KPKVK No 7.4/3-2023, 6541230, N.L.), MESU (No DI 247-22 M.P.), and
NFRU of Ukraine (No. 2020.02/0138 (M.O.T., N.V.V.).
\end{acknowledgments}

\bibliography{EE12370.bib}  %

\begin{thebibliography}{41}%
\makeatletter
\providecommand \@ifxundefined [1]{%
 \@ifx{#1\undefined}
}%
\providecommand \@ifnum [1]{%
 \ifnum #1\expandafter \@firstoftwo
 \else \expandafter \@secondoftwo
 \fi
}%
\providecommand \@ifx [1]{%
 \ifx #1\expandafter \@firstoftwo
 \else \expandafter \@secondoftwo
 \fi
}%
\providecommand \natexlab [1]{#1}%
\providecommand \enquote  [1]{``#1''}%
\providecommand \bibnamefont  [1]{#1}%
\providecommand \bibfnamefont [1]{#1}%
\providecommand \citenamefont [1]{#1}%
\providecommand \href@noop [0]{\@secondoftwo}%
\providecommand \href [0]{\begingroup \@sanitize@url \@href}%
\providecommand \@href[1]{\@@startlink{#1}\@@href}%
\providecommand \@@href[1]{\endgroup#1\@@endlink}%
\providecommand \@sanitize@url [0]{\catcode `\\12\catcode `\$12\catcode
  `\&12\catcode `\#12\catcode `\^12\catcode `\_12\catcode `\%12\relax}%
\providecommand \@@startlink[1]{}%
\providecommand \@@endlink[0]{}%
\providecommand \url  [0]{\begingroup\@sanitize@url \@url }%
\providecommand \@url [1]{\endgroup\@href {#1}{\urlprefix }}%
\providecommand \urlprefix  [0]{URL }%
\providecommand \Eprint [0]{\href }%
\providecommand \doibase [0]{https://doi.org/}%
\providecommand \selectlanguage [0]{\@gobble}%
\providecommand \bibinfo  [0]{\@secondoftwo}%
\providecommand \bibfield  [0]{\@secondoftwo}%
\providecommand \translation [1]{[#1]}%
\providecommand \BibitemOpen [0]{}%
\providecommand \bibitemStop [0]{}%
\providecommand \bibitemNoStop [0]{.\EOS\space}%
\providecommand \EOS [0]{\spacefactor3000\relax}%
\providecommand \BibitemShut  [1]{\csname bibitem#1\endcsname}%
\let\auto@bib@innerbib\@empty
\bibitem [{\citenamefont {Kubala}\ \emph {et~al.}(2022)\citenamefont {Kubala},
  \citenamefont {Batys}, \citenamefont {Barbasz}, \citenamefont {Weronski},\
  and\ \citenamefont {Ciesla}}]{Kubala2022}%
  \BibitemOpen
  \bibfield  {author} {\bibinfo {author} {\bibfnamefont {P.}~\bibnamefont
  {Kubala}}, \bibinfo {author} {\bibfnamefont {P.}~\bibnamefont {Batys}},
  \bibinfo {author} {\bibfnamefont {J.}~\bibnamefont {Barbasz}}, \bibinfo
  {author} {\bibfnamefont {P.}~\bibnamefont {Weronski}},\ and\ \bibinfo
  {author} {\bibfnamefont {M.}~\bibnamefont {Ciesla}},\ }\bibfield  {title}
  {\bibinfo {title} {Random sequential adsorption: An efficient tool for
  investigating the deposition of macromolecules and colloidal particles},\
  }\href {https://doi.org/10.1016/j.cis.2022.102692} {\bibfield  {journal}
  {\bibinfo  {journal} {Advances in Colloid and Interface Science}\ }\textbf
  {\bibinfo {volume} {306}},\ \bibinfo {pages} {102692} (\bibinfo {year}
  {2022})}\BibitemShut {NoStop}%
\bibitem [{\citenamefont {Adamczyk}\ \emph {et~al.}(2022)\citenamefont
  {Adamczyk}, \citenamefont {Morga}, \citenamefont {Nattich-Rak},\ and\
  \citenamefont {Sadowska}}]{Adamczyk2022}%
  \BibitemOpen
  \bibfield  {author} {\bibinfo {author} {\bibfnamefont {Z.}~\bibnamefont
  {Adamczyk}}, \bibinfo {author} {\bibfnamefont {M.}~\bibnamefont {Morga}},
  \bibinfo {author} {\bibfnamefont {M.}~\bibnamefont {Nattich-Rak}},\ and\
  \bibinfo {author} {\bibfnamefont {M.}~\bibnamefont {Sadowska}},\ }\bibfield
  {title} {\bibinfo {title} {Nanoparticle and bioparticle deposition
  kinetics},\ }\href {https://doi.org/10.1016/j.cis.2022.102630} {\bibfield
  {journal} {\bibinfo  {journal} {Advances in Colloid and Interface Science}\
  }\textbf {\bibinfo {volume} {302}},\ \bibinfo {pages} {102630} (\bibinfo
  {year} {2022})}\BibitemShut {NoStop}%
\bibitem [{\citenamefont {Evans}(1993)}]{Evans1993}%
  \BibitemOpen
  \bibfield  {author} {\bibinfo {author} {\bibfnamefont {J.~W.}\ \bibnamefont
  {Evans}},\ }\bibfield  {title} {\bibinfo {title} {Random and cooperative
  sequential adsorption},\ }\href {https://doi.org/10.1103/RevModPhys.65.1281}
  {\bibfield  {journal} {\bibinfo  {journal} {Reviews of Modern Physics}\
  }\textbf {\bibinfo {volume} {65}},\ \bibinfo {pages} {1281} (\bibinfo {year}
  {1993})}\BibitemShut {NoStop}%
\bibitem [{\citenamefont {Lebovka}\ and\ \citenamefont
  {Tarasevich}(2020)}]{Lebovka2020}%
  \BibitemOpen
  \bibfield  {author} {\bibinfo {author} {\bibfnamefont {N.~I.}\ \bibnamefont
  {Lebovka}}\ and\ \bibinfo {author} {\bibfnamefont {Y.~Y.}\ \bibnamefont
  {Tarasevich}},\ }\bibfield  {title} {\bibinfo {title} {Two-dimensional
  systems of elongated particles: From diluted to dense},\ }in\ \href
  {https://doi.org/10.1142/9789811216220\_0004} {\emph {\bibinfo {booktitle}
  {Order, Disorder and Criticality}}}\ (\bibinfo  {publisher} {World
  Scientific},\ \bibinfo {year} {2020})\ pp.\ \bibinfo {pages}
  {153--200}\BibitemShut {NoStop}%
\bibitem [{\citenamefont {Talbot}\ \emph {et~al.}(2000)\citenamefont {Talbot},
  \citenamefont {Tarjus}, \citenamefont {{Van Tassel}},\ and\ \citenamefont
  {Viot}}]{Talbot2000}%
  \BibitemOpen
  \bibfield  {author} {\bibinfo {author} {\bibfnamefont {J.}~\bibnamefont
  {Talbot}}, \bibinfo {author} {\bibfnamefont {G.}~\bibnamefont {Tarjus}},
  \bibinfo {author} {\bibfnamefont {P.}~\bibnamefont {{Van Tassel}}},\ and\
  \bibinfo {author} {\bibfnamefont {P.}~\bibnamefont {Viot}},\ }\bibfield
  {title} {\bibinfo {title} {From car parking to protein adsorption: an
  overview of sequential adsorption processes},\ }\href
  {https://doi.org/10.1016/S0927-7757(99)00409-4} {\bibfield  {journal}
  {\bibinfo  {journal} {Colloids and Surfaces A: Physicochemical and
  Engineering Aspects}\ }\textbf {\bibinfo {volume} {165}},\ \bibinfo {pages}
  {287} (\bibinfo {year} {2000})}\BibitemShut {NoStop}%
\bibitem [{\citenamefont {Feder}(1980)}]{Feder1980}%
  \BibitemOpen
  \bibfield  {author} {\bibinfo {author} {\bibfnamefont {J.}~\bibnamefont
  {Feder}},\ }\bibfield  {title} {\bibinfo {title} {Random sequential
  adsorption},\ }\href {https://doi.org/10.1016/0022-5193(80)90358-6}
  {\bibfield  {journal} {\bibinfo  {journal} {Journal of Theoretical Biology}\
  }\textbf {\bibinfo {volume} {87}},\ \bibinfo {pages} {237} (\bibinfo {year}
  {1980})}\BibitemShut {NoStop}%
\bibitem [{\citenamefont {Viot}\ and\ \citenamefont {Tarjus}(1990)}]{Viot1990}%
  \BibitemOpen
  \bibfield  {author} {\bibinfo {author} {\bibfnamefont {P.}~\bibnamefont
  {Viot}}\ and\ \bibinfo {author} {\bibfnamefont {G.}~\bibnamefont {Tarjus}},\
  }\bibfield  {title} {\bibinfo {title} {Random sequential addition of
  unoriented squares: Breakdown of swendsen's conjecture},\ }\href
  {https://doi.org/10.1209/0295-5075/13/4/002} {\bibfield  {journal} {\bibinfo
  {journal} {Europhysics Letters (EPL)}\ }\textbf {\bibinfo {volume} {13}},\
  \bibinfo {pages} {295} (\bibinfo {year} {1990})}\BibitemShut {NoStop}%
\bibitem [{\citenamefont {Malmir}\ \emph {et~al.}(2016)\citenamefont {Malmir},
  \citenamefont {Sahimi},\ and\ \citenamefont {Tabar}}]{Malmir2016}%
  \BibitemOpen
  \bibfield  {author} {\bibinfo {author} {\bibfnamefont {H.}~\bibnamefont
  {Malmir}}, \bibinfo {author} {\bibfnamefont {M.}~\bibnamefont {Sahimi}},\
  and\ \bibinfo {author} {\bibfnamefont {M.~R.~R.}\ \bibnamefont {Tabar}},\
  }\bibfield  {title} {\bibinfo {title} {Packing of nonoverlapping cubic
  particles: Computational algorithms and microstructural characteristics},\
  }\href {https://doi.org/10.1103/PhysRevE.94.062901} {\bibfield  {journal}
  {\bibinfo  {journal} {Physical Review E}\ }\textbf {\bibinfo {volume} {94}},\
  \bibinfo {pages} {062901} (\bibinfo {year} {2016})}\BibitemShut {NoStop}%
\bibitem [{\citenamefont {Vigil}\ and\ \citenamefont {Ziff}(1989)}]{Vigil1989}%
  \BibitemOpen
  \bibfield  {author} {\bibinfo {author} {\bibfnamefont {R.~D.}\ \bibnamefont
  {Vigil}}\ and\ \bibinfo {author} {\bibfnamefont {R.~M.}\ \bibnamefont
  {Ziff}},\ }\bibfield  {title} {\bibinfo {title} {Random sequential adsorption
  of unoriented rectangles onto a plane},\ }\href
  {https://doi.org/10.1063/1.457021} {\bibfield  {journal} {\bibinfo  {journal}
  {The Journal of Chemical Physics}\ }\textbf {\bibinfo {volume} {91}},\
  \bibinfo {pages} {2599} (\bibinfo {year} {1989})}\BibitemShut {NoStop}%
\bibitem [{\citenamefont {Vigil}\ and\ \citenamefont {Ziff}(1990)}]{Vigil1990}%
  \BibitemOpen
  \bibfield  {author} {\bibinfo {author} {\bibfnamefont {R.~D.}\ \bibnamefont
  {Vigil}}\ and\ \bibinfo {author} {\bibfnamefont {R.~M.}\ \bibnamefont
  {Ziff}},\ }\bibfield  {title} {\bibinfo {title} {Kinetics of random
  sequential adsorption of rectangles and line segments},\ }\href
  {https://doi.org/10.1063/1.459307} {\bibfield  {journal} {\bibinfo  {journal}
  {The Journal of Chemical Physics}\ }\textbf {\bibinfo {volume} {93}},\
  \bibinfo {pages} {8270} (\bibinfo {year} {1990})}\BibitemShut {NoStop}%
\bibitem [{\citenamefont {Viot}\ \emph {et~al.}(1992)\citenamefont {Viot},
  \citenamefont {Tarjus}, \citenamefont {Ricci},\ and\ \citenamefont
  {Talbot}}]{Viot1992}%
  \BibitemOpen
  \bibfield  {author} {\bibinfo {author} {\bibfnamefont {P.}~\bibnamefont
  {Viot}}, \bibinfo {author} {\bibfnamefont {G.}~\bibnamefont {Tarjus}},
  \bibinfo {author} {\bibfnamefont {S.~M.}\ \bibnamefont {Ricci}},\ and\
  \bibinfo {author} {\bibfnamefont {J.}~\bibnamefont {Talbot}},\ }\bibfield
  {title} {\bibinfo {title} {Random sequential adsorption of anisotropic
  particles. i. jamming limit and asymptotic behavior},\ }\href
  {https://doi.org/10.1063/1.463820} {\bibfield  {journal} {\bibinfo  {journal}
  {The Journal of Chemical Physics}\ }\textbf {\bibinfo {volume} {97}},\
  \bibinfo {pages} {5212} (\bibinfo {year} {1992})}\BibitemShut {NoStop}%
\bibitem [{\citenamefont {Ricci}\ \emph {et~al.}(1992)\citenamefont {Ricci},
  \citenamefont {Talbot}, \citenamefont {Tarjus},\ and\ \citenamefont
  {Viot}}]{Ricci1992}%
  \BibitemOpen
  \bibfield  {author} {\bibinfo {author} {\bibfnamefont {S.~M.}\ \bibnamefont
  {Ricci}}, \bibinfo {author} {\bibfnamefont {J.}~\bibnamefont {Talbot}},
  \bibinfo {author} {\bibfnamefont {G.}~\bibnamefont {Tarjus}},\ and\ \bibinfo
  {author} {\bibfnamefont {P.}~\bibnamefont {Viot}},\ }\bibfield  {title}
  {\bibinfo {title} {Random sequential adsorption of anisotropic particles. ii.
  low coverage kinetics},\ }\href {https://doi.org/10.1063/1.463988} {\bibfield
   {journal} {\bibinfo  {journal} {The Journal of Chemical Physics}\ }\textbf
  {\bibinfo {volume} {97}},\ \bibinfo {pages} {5219} (\bibinfo {year}
  {1992})}\BibitemShut {NoStop}%
\bibitem [{\citenamefont {Kasperek}\ \emph {et~al.}(2018)\citenamefont
  {Kasperek}, \citenamefont {Kubala},\ and\ \citenamefont
  {Cie\ifmmode~\acute{s}\else \'{s}\fi{}la}}]{Kasperek2018}%
  \BibitemOpen
  \bibfield  {author} {\bibinfo {author} {\bibfnamefont {W.}~\bibnamefont
  {Kasperek}}, \bibinfo {author} {\bibfnamefont {P.}~\bibnamefont {Kubala}},\
  and\ \bibinfo {author} {\bibfnamefont {M.}~\bibnamefont
  {Cie\ifmmode~\acute{s}\else \'{s}\fi{}la}},\ }\bibfield  {title} {\bibinfo
  {title} {Random sequential adsorption of unoriented rectangles at
  saturation},\ }\href {https://doi.org/10.1103/PhysRevE.98.063310} {\bibfield
  {journal} {\bibinfo  {journal} {Physical Review E}\ }\textbf {\bibinfo
  {volume} {98}},\ \bibinfo {pages} {063310} (\bibinfo {year}
  {2018})}\BibitemShut {NoStop}%
\bibitem [{\citenamefont {Petrone}\ and\ \citenamefont
  {Cie\ifmmode~\acute{s}\else \'{s}\fi{}la}(2021)}]{Petrone2021}%
  \BibitemOpen
  \bibfield  {author} {\bibinfo {author} {\bibfnamefont {L.}~\bibnamefont
  {Petrone}}\ and\ \bibinfo {author} {\bibfnamefont {M.}~\bibnamefont
  {Cie\ifmmode~\acute{s}\else \'{s}\fi{}la}},\ }\bibfield  {title} {\bibinfo
  {title} {Random sequential adsorption of oriented rectangles with random
  aspect ratio},\ }\href {https://doi.org/10.1103/PhysRevE.104.034903}
  {\bibfield  {journal} {\bibinfo  {journal} {Physical Review E}\ }\textbf
  {\bibinfo {volume} {104}},\ \bibinfo {pages} {034903} (\bibinfo {year}
  {2021})}\BibitemShut {NoStop}%
\bibitem [{\citenamefont {Haiduk}\ \emph {et~al.}(2018)\citenamefont {Haiduk},
  \citenamefont {Kubala},\ and\ \citenamefont {Cie\'{s}la}}]{Haiduk2018}%
  \BibitemOpen
  \bibfield  {author} {\bibinfo {author} {\bibfnamefont {K.}~\bibnamefont
  {Haiduk}}, \bibinfo {author} {\bibfnamefont {P.}~\bibnamefont {Kubala}},\
  and\ \bibinfo {author} {\bibfnamefont {M.}~\bibnamefont {Cie\'{s}la}},\
  }\bibfield  {title} {\bibinfo {title} {Saturated packings of convex
  anisotropic objects under random sequential adsorption protocol},\ }\href
  {https://doi.org/10.1103/PhysRevE.98.063309} {\bibfield  {journal} {\bibinfo
  {journal} {Physical Review E}\ }\textbf {\bibinfo {volume} {98}},\ \bibinfo
  {pages} {063309} (\bibinfo {year} {2018})}\BibitemShut {NoStop}%
\bibitem [{\citenamefont {Cieśla}\ \emph {et~al.}(2020)\citenamefont
  {Cieśla}, \citenamefont {Kozubek}, \citenamefont {Kubala},\ and\
  \citenamefont {Baule}}]{Ciesla2020}%
  \BibitemOpen
  \bibfield  {author} {\bibinfo {author} {\bibfnamefont {M.}~\bibnamefont
  {Cieśla}}, \bibinfo {author} {\bibfnamefont {K.}~\bibnamefont {Kozubek}},
  \bibinfo {author} {\bibfnamefont {P.}~\bibnamefont {Kubala}},\ and\ \bibinfo
  {author} {\bibfnamefont {A.}~\bibnamefont {Baule}},\ }\bibfield  {title}
  {\bibinfo {title} {Kinetics of random sequential adsorption of
  two-dimensional shapes on a one-dimensional line},\ }\href
  {https://doi.org/10.1103/PhysRevE.101.042901} {\bibfield  {journal} {\bibinfo
   {journal} {Physical Review E}\ }\textbf {\bibinfo {volume} {101}},\ \bibinfo
  {pages} {042901} (\bibinfo {year} {2020})}\BibitemShut {NoStop}%
\bibitem [{\citenamefont {Talbot}\ and\ \citenamefont
  {Schaaf}(1989)}]{Talbot1989}%
  \BibitemOpen
  \bibfield  {author} {\bibinfo {author} {\bibfnamefont {J.}~\bibnamefont
  {Talbot}}\ and\ \bibinfo {author} {\bibfnamefont {P.}~\bibnamefont
  {Schaaf}},\ }\bibfield  {title} {\bibinfo {title} {Random sequential
  adsorption of mixtures},\ }\href {https://doi.org/10.1103/PhysRevA.40.422}
  {\bibfield  {journal} {\bibinfo  {journal} {Physical Review A}\ }\textbf
  {\bibinfo {volume} {40}},\ \bibinfo {pages} {422} (\bibinfo {year}
  {1989})}\BibitemShut {NoStop}%
\bibitem [{\citenamefont {Sherwood}(1990)}]{Sherwood1990}%
  \BibitemOpen
  \bibfield  {author} {\bibinfo {author} {\bibfnamefont {J.~D.}\ \bibnamefont
  {Sherwood}},\ }\bibfield  {title} {\bibinfo {title} {Random sequential
  adsorption of lines and ellipses},\ }\href
  {https://doi.org/10.1088/0305-4470/23/13/021} {\bibfield  {journal} {\bibinfo
   {journal} {Journal of Physics A: Mathematical and General}\ }\textbf
  {\bibinfo {volume} {23}},\ \bibinfo {pages} {2827} (\bibinfo {year}
  {1990})}\BibitemShut {NoStop}%
\bibitem [{\citenamefont {Zhang}(2018)}]{Zhang2018}%
  \BibitemOpen
  \bibfield  {author} {\bibinfo {author} {\bibfnamefont {G.}~\bibnamefont
  {Zhang}},\ }\bibfield  {title} {\bibinfo {title} {Precise algorithm to
  generate random sequential adsorption of hard polygons at saturation},\
  }\href {https://doi.org/10.1103/PhysRevE.97.043311} {\bibfield  {journal}
  {\bibinfo  {journal} {Physical Review E}\ }\textbf {\bibinfo {volume} {97}},\
  \bibinfo {pages} {043311} (\bibinfo {year} {2018})}\BibitemShut {NoStop}%
\bibitem [{\citenamefont {Morga}\ \emph {et~al.}(2022)\citenamefont {Morga},
  \citenamefont {Nattich-Rak}, \citenamefont {Adamczyk}, \citenamefont
  {Mickiewicz}, \citenamefont {Gadzinowski},\ and\ \citenamefont
  {Basinska}}]{Morga2022}%
  \BibitemOpen
  \bibfield  {author} {\bibinfo {author} {\bibfnamefont {M.}~\bibnamefont
  {Morga}}, \bibinfo {author} {\bibfnamefont {M.}~\bibnamefont {Nattich-Rak}},
  \bibinfo {author} {\bibfnamefont {Z.}~\bibnamefont {Adamczyk}}, \bibinfo
  {author} {\bibfnamefont {D.}~\bibnamefont {Mickiewicz}}, \bibinfo {author}
  {\bibfnamefont {M.}~\bibnamefont {Gadzinowski}},\ and\ \bibinfo {author}
  {\bibfnamefont {T.}~\bibnamefont {Basinska}},\ }\bibfield  {title} {\bibinfo
  {title} {Mechanisms of anisotropic particle deposition: Prolate spheroid
  layers on mica},\ }\href {https://doi.org/10.1021/acs.jpcc.2c06028}
  {\bibfield  {journal} {\bibinfo  {journal} {The Journal of Physical Chemistry
  C}\ }\textbf {\bibinfo {volume} {126}},\ \bibinfo {pages} {18550} (\bibinfo
  {year} {2022})}\BibitemShut {NoStop}%
\bibitem [{\citenamefont {Tarjus}\ and\ \citenamefont
  {Viot}(1991)}]{Tarjus1991}%
  \BibitemOpen
  \bibfield  {author} {\bibinfo {author} {\bibfnamefont {G.}~\bibnamefont
  {Tarjus}}\ and\ \bibinfo {author} {\bibfnamefont {P.}~\bibnamefont {Viot}},\
  }\bibfield  {title} {\bibinfo {title} {Asymptotic results for the random
  sequential addition of unoriented objects},\ }\href
  {https://doi.org/10.1103/PhysRevLett.67.1875} {\bibfield  {journal} {\bibinfo
   {journal} {Physical Review Letters}\ }\textbf {\bibinfo {volume} {67}},\
  \bibinfo {pages} {1875} (\bibinfo {year} {1991})}\BibitemShut {NoStop}%
\bibitem [{\citenamefont {Chaikin}\ \emph {et~al.}(2006)\citenamefont
  {Chaikin}, \citenamefont {Donev}, \citenamefont {Man}, \citenamefont
  {Stillinger},\ and\ \citenamefont {Torquato}}]{Chaikin2006}%
  \BibitemOpen
  \bibfield  {author} {\bibinfo {author} {\bibfnamefont {P.~M.}\ \bibnamefont
  {Chaikin}}, \bibinfo {author} {\bibfnamefont {A.}~\bibnamefont {Donev}},
  \bibinfo {author} {\bibfnamefont {W.}~\bibnamefont {Man}}, \bibinfo {author}
  {\bibfnamefont {F.~H.}\ \bibnamefont {Stillinger}},\ and\ \bibinfo {author}
  {\bibfnamefont {S.}~\bibnamefont {Torquato}},\ }\bibfield  {title} {\bibinfo
  {title} {Some observations on the random packing of hard ellipsoids},\ }\href
  {https://doi.org/10.1021/ie060032g} {\bibfield  {journal} {\bibinfo
  {journal} {Industrial \& engineering chemistry research}\ }\textbf {\bibinfo
  {volume} {45}},\ \bibinfo {pages} {6960} (\bibinfo {year}
  {2006})}\BibitemShut {NoStop}%
\bibitem [{\citenamefont {Meakin}\ and\ \citenamefont
  {Jullien}(1992)}]{Meakin1992}%
  \BibitemOpen
  \bibfield  {author} {\bibinfo {author} {\bibfnamefont {P.}~\bibnamefont
  {Meakin}}\ and\ \bibinfo {author} {\bibfnamefont {R.}~\bibnamefont
  {Jullien}},\ }\bibfield  {title} {\bibinfo {title} {Random-sequential
  adsorption of disks of different sizes},\ }\href
  {https://doi.org/10.1103/PhysRevA.46.2029} {\bibfield  {journal} {\bibinfo
  {journal} {Physical Review A}\ }\textbf {\bibinfo {volume} {46}},\ \bibinfo
  {pages} {2029} (\bibinfo {year} {1992})}\BibitemShut {NoStop}%
\bibitem [{\citenamefont {Wagaskar}\ \emph {et~al.}(2020)\citenamefont
  {Wagaskar}, \citenamefont {Late}, \citenamefont {Banpurkar}, \citenamefont
  {Limaye},\ and\ \citenamefont {Shelke}}]{Wagaskar2020}%
  \BibitemOpen
  \bibfield  {author} {\bibinfo {author} {\bibfnamefont {K.~V.}\ \bibnamefont
  {Wagaskar}}, \bibinfo {author} {\bibfnamefont {R.}~\bibnamefont {Late}},
  \bibinfo {author} {\bibfnamefont {A.~G.}\ \bibnamefont {Banpurkar}}, \bibinfo
  {author} {\bibfnamefont {A.~V.}\ \bibnamefont {Limaye}},\ and\ \bibinfo
  {author} {\bibfnamefont {P.~B.}\ \bibnamefont {Shelke}},\ }\bibfield  {title}
  {\bibinfo {title} {Simulation studies of random sequential adsorption ({RSA})
  of mixture of two-component circular discs},\ }\href
  {https://doi.org/10.1007/s10955-020-02660-7} {\bibfield  {journal} {\bibinfo
  {journal} {Journal of Statistical Physics}\ }\textbf {\bibinfo {volume}
  {181}},\ \bibinfo {pages} {2191} (\bibinfo {year} {2020})}\BibitemShut
  {NoStop}%
\bibitem [{\citenamefont {Martins}\ \emph {et~al.}(2023)\citenamefont
  {Martins}, \citenamefont {Dickman},\ and\ \citenamefont
  {Ziff}}]{Martins2023}%
  \BibitemOpen
  \bibfield  {author} {\bibinfo {author} {\bibfnamefont {P.~H.~L.}\
  \bibnamefont {Martins}}, \bibinfo {author} {\bibfnamefont {R.}~\bibnamefont
  {Dickman}},\ and\ \bibinfo {author} {\bibfnamefont {R.~M.}\ \bibnamefont
  {Ziff}},\ }\bibfield  {title} {\bibinfo {title} {Percolation in two-species
  antagonistic random sequential adsorption in two dimensions},\ }\href
  {https://doi.org/10.1103/PhysRevE.107.024104} {\bibfield  {journal} {\bibinfo
   {journal} {Physical Review E}\ }\textbf {\bibinfo {volume} {107}},\ \bibinfo
  {pages} {024104} (\bibinfo {year} {2023})}\BibitemShut {NoStop}%
\bibitem [{\citenamefont {Ara\'ujo}\ \emph {et~al.}(2008)\citenamefont
  {Ara\'ujo}, \citenamefont {Cadilhe},\ and\ \citenamefont
  {Privman}}]{Araujo2008}%
  \BibitemOpen
  \bibfield  {author} {\bibinfo {author} {\bibfnamefont {N.~A.~M.}\
  \bibnamefont {Ara\'ujo}}, \bibinfo {author} {\bibfnamefont {A.}~\bibnamefont
  {Cadilhe}},\ and\ \bibinfo {author} {\bibfnamefont {V.}~\bibnamefont
  {Privman}},\ }\bibfield  {title} {\bibinfo {title} {Morphology of
  fine-particle monolayers deposited on nanopatterned substrates},\ }\href
  {https://doi.org/10.1103/PhysRevE.77.031603} {\bibfield  {journal} {\bibinfo
  {journal} {Physical Review E}\ }\textbf {\bibinfo {volume} {77}},\ \bibinfo
  {pages} {031603} (\bibinfo {year} {2008})}\BibitemShut {NoStop}%
\bibitem [{\citenamefont {Stojiljković}\ \emph {et~al.}(2015)\citenamefont
  {Stojiljković}, \citenamefont {Šćepanović}, \citenamefont {Vrhovac},\
  and\ \citenamefont {Švrakić}}]{Stojiljkovic2015}%
  \BibitemOpen
  \bibfield  {author} {\bibinfo {author} {\bibfnamefont {D.}~\bibnamefont
  {Stojiljković}}, \bibinfo {author} {\bibfnamefont {J.}~\bibnamefont
  {Šćepanović}}, \bibinfo {author} {\bibfnamefont {S.}~\bibnamefont
  {Vrhovac}},\ and\ \bibinfo {author} {\bibfnamefont {N.}~\bibnamefont
  {Švrakić}},\ }\bibfield  {title} {\bibinfo {title} {Structural properties
  of particle deposits at heterogeneous surfaces},\ }\href
  {https://doi.org/10.1088/1742-5468/2015/06/P06032} {\bibfield  {journal}
  {\bibinfo  {journal} {Journal of Statistical Mechanics: Theory and
  Experiment}\ }\textbf {\bibinfo {volume} {2015}},\ \bibinfo {pages} {P06032}
  (\bibinfo {year} {2015})}\BibitemShut {NoStop}%
\bibitem [{\citenamefont {Adamczyk}\ \emph {et~al.}(1997)\citenamefont
  {Adamczyk}, \citenamefont {Siwek}, \citenamefont {Zembala},\ and\
  \citenamefont {Weroński}}]{Adamczyk1997}%
  \BibitemOpen
  \bibfield  {author} {\bibinfo {author} {\bibfnamefont {Z.}~\bibnamefont
  {Adamczyk}}, \bibinfo {author} {\bibfnamefont {B.}~\bibnamefont {Siwek}},
  \bibinfo {author} {\bibfnamefont {M.}~\bibnamefont {Zembala}},\ and\ \bibinfo
  {author} {\bibfnamefont {P.}~\bibnamefont {Weroński}},\ }\bibfield  {title}
  {\bibinfo {title} {Influence of polydispersity on random sequential
  adsorption of spherical particles},\ }\href
  {https://doi.org/https://doi.org/10.1006/jcis.1996.4540} {\bibfield
  {journal} {\bibinfo  {journal} {Journal of Colloid and Interface Science}\
  }\textbf {\bibinfo {volume} {185}},\ \bibinfo {pages} {236} (\bibinfo {year}
  {1997})}\BibitemShut {NoStop}%
\bibitem [{\citenamefont {Adamczyk}\ \emph {et~al.}(1998)\citenamefont
  {Adamczyk}, \citenamefont {Siwek}, \citenamefont {Weronski},\ and\
  \citenamefont {Zembala}}]{Adamczyk1998}%
  \BibitemOpen
  \bibfield  {author} {\bibinfo {author} {\bibfnamefont {Z.}~\bibnamefont
  {Adamczyk}}, \bibinfo {author} {\bibfnamefont {B.}~\bibnamefont {Siwek}},
  \bibinfo {author} {\bibfnamefont {P.}~\bibnamefont {Weronski}},\ and\
  \bibinfo {author} {\bibfnamefont {M.}~\bibnamefont {Zembala}},\ }\bibfield
  {title} {\bibinfo {title} {Adsorption of colloid particle mixtures at
  interfaces},\ }in\ \href {https://doi.org/10.1007/BFb0118107} {\emph
  {\bibinfo {booktitle} {Structure, Dynamics and Properties of Disperse
  Colloidal Systems}}}\ (\bibinfo  {publisher} {Steinkopff},\ \bibinfo
  {address} {Darmstadt},\ \bibinfo {year} {1998})\ pp.\ \bibinfo {pages}
  {41--47}\BibitemShut {NoStop}%
\bibitem [{\citenamefont {Weroński}(2005)}]{Weronski2005}%
  \BibitemOpen
  \bibfield  {author} {\bibinfo {author} {\bibfnamefont {P.}~\bibnamefont
  {Weroński}},\ }\bibfield  {title} {\bibinfo {title} {Application of the
  extended rsa models in studies of particle deposition at partially covered
  surfaces},\ }\href {https://doi.org/10.1016/j.cis.2005.03.002} {\bibfield
  {journal} {\bibinfo  {journal} {Advances in Colloid and Interface Science}\
  }\textbf {\bibinfo {volume} {118}},\ \bibinfo {pages} {1} (\bibinfo {year}
  {2005})}\BibitemShut {NoStop}%
\bibitem [{\citenamefont {Manciu}\ and\ \citenamefont
  {Ruckenstein}(2004)}]{Manciu2004}%
  \BibitemOpen
  \bibfield  {author} {\bibinfo {author} {\bibfnamefont {M.}~\bibnamefont
  {Manciu}}\ and\ \bibinfo {author} {\bibfnamefont {E.}~\bibnamefont
  {Ruckenstein}},\ }\bibfield  {title} {\bibinfo {title} {Estimation of the
  available surface and the jamming coverage in the random sequential
  adsorption of a binary mixture of disks},\ }\href
  {https://doi.org/10.1016/j.colsurfa.2003.10.001} {\bibfield  {journal}
  {\bibinfo  {journal} {Colloids and Surfaces A: Physicochemical and
  Engineering Aspects}\ }\textbf {\bibinfo {volume} {232}},\ \bibinfo {pages}
  {1} (\bibinfo {year} {2004})}\BibitemShut {NoStop}%
\bibitem [{\citenamefont {Adamczyk}\ \emph {et~al.}(2002)\citenamefont
  {Adamczyk}, \citenamefont {Siwek}, \citenamefont {Weroński},\ and\
  \citenamefont {Musia{\l}}}]{Adamczyk2002}%
  \BibitemOpen
  \bibfield  {author} {\bibinfo {author} {\bibfnamefont {Z.}~\bibnamefont
  {Adamczyk}}, \bibinfo {author} {\bibfnamefont {B.}~\bibnamefont {Siwek}},
  \bibinfo {author} {\bibfnamefont {P.}~\bibnamefont {Weroński}},\ and\
  \bibinfo {author} {\bibfnamefont {E.}~\bibnamefont {Musia{\l}}},\ }\bibfield
  {title} {\bibinfo {title} {Irreversible adsorption of colloid particles at
  heterogeneous surfaces},\ }\href
  {https://doi.org/10.1016/S0169-4332(02)00063-6} {\bibfield  {journal}
  {\bibinfo  {journal} {Applied Surface Science}\ }\textbf {\bibinfo {volume}
  {196}},\ \bibinfo {pages} {250} (\bibinfo {year} {2002})}\BibitemShut
  {NoStop}%
\bibitem [{\citenamefont {Weroński}(2007{\natexlab{a}})}]{Weronski2007}%
  \BibitemOpen
  \bibfield  {author} {\bibinfo {author} {\bibfnamefont {P.}~\bibnamefont
  {Weroński}},\ }\bibfield  {title} {\bibinfo {title} {Effect of electrostatic
  interaction on deposition of colloid on partially covered surfaces: Part i.
  model formulation},\ }\href {https://doi.org/10.1016/j.colsurfa.2006.08.018}
  {\bibfield  {journal} {\bibinfo  {journal} {Colloids and Surfaces A:
  Physicochemical and Engineering Aspects}\ }\textbf {\bibinfo {volume}
  {294}},\ \bibinfo {pages} {254} (\bibinfo {year}
  {2007}{\natexlab{a}})}\BibitemShut {NoStop}%
\bibitem [{\citenamefont {Weroński}(2007{\natexlab{b}})}]{Weronski2007a}%
  \BibitemOpen
  \bibfield  {author} {\bibinfo {author} {\bibfnamefont {P.}~\bibnamefont
  {Weroński}},\ }\bibfield  {title} {\bibinfo {title} {Effect of electrostatic
  interaction on deposition of colloid on partially covered surfaces: Part ii.
  results of computer simulations},\ }\href
  {https://doi.org/10.1016/j.colsurfa.2006.08.020} {\bibfield  {journal}
  {\bibinfo  {journal} {Colloids and Surfaces A: Physicochemical and
  Engineering Aspects}\ }\textbf {\bibinfo {volume} {294}},\ \bibinfo {pages}
  {267} (\bibinfo {year} {2007}{\natexlab{b}})}\BibitemShut {NoStop}%
\bibitem [{\citenamefont {Sadowska}\ \emph {et~al.}(2021)\citenamefont
  {Sadowska}, \citenamefont {Cieśla},\ and\ \citenamefont
  {Adamczyk}}]{Sadowska2021}%
  \BibitemOpen
  \bibfield  {author} {\bibinfo {author} {\bibfnamefont {M.}~\bibnamefont
  {Sadowska}}, \bibinfo {author} {\bibfnamefont {M.}~\bibnamefont {Cieśla}},\
  and\ \bibinfo {author} {\bibfnamefont {Z.}~\bibnamefont {Adamczyk}},\
  }\bibfield  {title} {\bibinfo {title} {Nanoparticle deposition on
  heterogeneous surfaces: Random sequential adsorption modeling and
  experiments},\ }\href {https://doi.org/10.1016/j.colsurfa.2021.126296}
  {\bibfield  {journal} {\bibinfo  {journal} {Colloids and Surfaces A:
  Physicochemical and Engineering Aspects}\ }\textbf {\bibinfo {volume}
  {617}},\ \bibinfo {pages} {126296} (\bibinfo {year} {2021})}\BibitemShut
  {NoStop}%
\bibitem [{\citenamefont {Adamczyk}(2017)}]{Adamczyk2017}%
  \BibitemOpen
  \bibfield  {author} {\bibinfo {author} {\bibfnamefont {Z.}~\bibnamefont
  {Adamczyk}},\ }\href@noop {} {\emph {\bibinfo {title} {Particles at
  interfaces: Interactions, deposition, structure}}}\ (\bibinfo  {publisher}
  {Academic Press},\ \bibinfo {year} {2017})\BibitemShut {NoStop}%
\bibitem [{\citenamefont {Lebovka}\ \emph {et~al.}(2021)\citenamefont
  {Lebovka}, \citenamefont {Tatochenko}, \citenamefont {Vygornitskii},
  \citenamefont {Eserkepov}, \citenamefont {Akhunzhanov},\ and\ \citenamefont
  {Tarasevich}}]{Lebovka2021}%
  \BibitemOpen
  \bibfield  {author} {\bibinfo {author} {\bibfnamefont {N.~I.}\ \bibnamefont
  {Lebovka}}, \bibinfo {author} {\bibfnamefont {M.~O.}\ \bibnamefont
  {Tatochenko}}, \bibinfo {author} {\bibfnamefont {N.~V.}\ \bibnamefont
  {Vygornitskii}}, \bibinfo {author} {\bibfnamefont {A.~V.}\ \bibnamefont
  {Eserkepov}}, \bibinfo {author} {\bibfnamefont {R.~K.}\ \bibnamefont
  {Akhunzhanov}},\ and\ \bibinfo {author} {\bibfnamefont {Y.~Y.}\ \bibnamefont
  {Tarasevich}},\ }\bibfield  {title} {\bibinfo {title} {Connectedness
  percolation in the random sequential adsorption packings of elongated
  particles},\ }\href {https://doi.org/10.1103/PhysRevE.103.042113} {\bibfield
  {journal} {\bibinfo  {journal} {Physical Review E}\ }\textbf {\bibinfo
  {volume} {103}},\ \bibinfo {pages} {042113} (\bibinfo {year}
  {2021})}\BibitemShut {NoStop}%
\bibitem [{\citenamefont {van~der Marck}(1997)}]{Marck1997}%
  \BibitemOpen
  \bibfield  {author} {\bibinfo {author} {\bibfnamefont {S.~C.}\ \bibnamefont
  {van~der Marck}},\ }\bibfield  {title} {\bibinfo {title} {Percolation
  thresholds and universal formulas},\ }\href
  {https://doi.org/10.1103/PhysRevE.55.1514} {\bibfield  {journal} {\bibinfo
  {journal} {Physical Review E}\ }\textbf {\bibinfo {volume} {55}},\ \bibinfo
  {pages} {1514} (\bibinfo {year} {1997})}\BibitemShut {NoStop}%
\bibitem [{\citenamefont {Hoshen}\ and\ \citenamefont
  {Kopelman}(1976)}]{Hoshen1976}%
  \BibitemOpen
  \bibfield  {author} {\bibinfo {author} {\bibfnamefont {J.}~\bibnamefont
  {Hoshen}}\ and\ \bibinfo {author} {\bibfnamefont {R.}~\bibnamefont
  {Kopelman}},\ }\bibfield  {title} {\bibinfo {title} {Percolation and cluster
  distribution. i. cluster multiple labeling technique and critical
  concentration algorithm},\ }\href {https://doi.org/10.1103/PhysRevB.14.3438}
  {\bibfield  {journal} {\bibinfo  {journal} {Physical Review B}\ }\textbf
  {\bibinfo {volume} {14}},\ \bibinfo {pages} {3438} (\bibinfo {year}
  {1976})}\BibitemShut {NoStop}%
\bibitem [{\citenamefont {Hinrichsen}\ \emph {et~al.}(1986)\citenamefont
  {Hinrichsen}, \citenamefont {Feder},\ and\ \citenamefont
  {Jøssang}}]{Hinrichsen1986}%
  \BibitemOpen
  \bibfield  {author} {\bibinfo {author} {\bibfnamefont {E.~L.}\ \bibnamefont
  {Hinrichsen}}, \bibinfo {author} {\bibfnamefont {J.}~\bibnamefont {Feder}},\
  and\ \bibinfo {author} {\bibfnamefont {T.}~\bibnamefont {Jøssang}},\
  }\bibfield  {title} {\bibinfo {title} {Geometry of random sequential
  adsorption},\ }\href {https://doi.org/10.1007/BF01011908} {\bibfield
  {journal} {\bibinfo  {journal} {Journal of Statistical Physics}\ }\textbf
  {\bibinfo {volume} {44}},\ \bibinfo {pages} {793} (\bibinfo {year}
  {1986})}\BibitemShut {NoStop}%
\bibitem [{\citenamefont {Quintanilla}\ \emph {et~al.}(2000)\citenamefont
  {Quintanilla}, \citenamefont {Torquato},\ and\ \citenamefont
  {Ziff}}]{Quintanilla2000}%
  \BibitemOpen
  \bibfield  {author} {\bibinfo {author} {\bibfnamefont {J.}~\bibnamefont
  {Quintanilla}}, \bibinfo {author} {\bibfnamefont {S.}~\bibnamefont
  {Torquato}},\ and\ \bibinfo {author} {\bibfnamefont {R.~M.}\ \bibnamefont
  {Ziff}},\ }\bibfield  {title} {\bibinfo {title} {Efficient measurement of the
  percolation threshold for fully penetrable discs},\ }\href
  {https://doi.org/10.1088/0305-4470/33/42/104} {\bibfield  {journal} {\bibinfo
   {journal} {Journal of Physics A: Mathematical and General}\ }\textbf
  {\bibinfo {volume} {33}},\ \bibinfo {pages} {L399} (\bibinfo {year}
  {2000})}\BibitemShut {NoStop}%
\end{thebibliography}%

\end{document}